\documentclass{article}
\usepackage{afterpage}
\usepackage{longtable}
\usepackage{graphicx}
\usepackage{enumitem}
\usepackage{hyperref}

   \usepackage{supertabular}

\usepackage{array}
\usepackage{makecell}
\usepackage{tabularx}

\usepackage{blindtext}

\usepackage{rotating}
\usepackage{colortbl}
\usepackage{multirow}
\usepackage{makecell}
\newcommand{\myrotcell}[1]{\rotcell{\makebox[0pt][c]{#1}}}

\usepackage{tabularx} 

\usepackage{placeins}

\usepackage{etoolbox}
\AtBeginEnvironment{longtable}{%
   \setlength\thickmuskip{0mu}%
   \setlength\tabcolsep{4pt}
   \setlength\LTcapwidth{\textwidth}%
   }

\AtBeginDocument{%
  \providecommand\BibTeX{{%
    \normalfont B\kern-0.5em{\scshape i\kern-0.25em b}\kern-0.8em\TeX}}}

\usepackage{graphicx}
\graphicspath{ {./figures/} }

\usepackage[T1]{fontenc}

\usepackage{amsmath}

\usepackage[cmintegrals]{newtxmath}

\usepackage{arxiv}

\usepackage[utf8]{inputenc} 
\usepackage[T1]{fontenc}    
\usepackage{hyperref}       
\usepackage{url}            
\usepackage{booktabs}       
\usepackage{amsfonts}       

\usepackage{microtype}      
\usepackage{lipsum}
\usepackage{graphicx}
\usepackage{longtable}
\usepackage{rotating}

\usepackage{tikz}

\usepackage{colortbl}

\graphicspath{ {./images/} }

\title{Privacy-Aware Internet of Things Notices in Shared Spaces: A Survey}

\author{
 Bayan~Al Muhander \\
  School of Computer Science and Informatics\\
  Cardiff University, UK \\
  \texttt{almuhanderb@cardiff.ac.uk} \\
   \And
    Jason~Wiese \\
  School of Computing\\
  University of Utah, USA \\
  \texttt{wiese@cs.utah.edu} \\
  \And
    Omer~Rana \\
  School of Computer Science and Informatics\\
  Cardiff University, UK \\
  \texttt{RanaOF@cardiff.ac.uk} \\
  \And
  Charith~Perera \\
  School of Computer Science and Informatics\\
  Cardiff University, UK \\
  \texttt{pererac@cardiff.ac.uk} \\
}
\usepackage{subcaption}

\begin{document}
\maketitle
\begin{abstract}
The balance between protecting users' privacy while providing cost-effective devices that are functional and usable is a key challenge in the burgeoning Internet of Things (IoT) industry. While in traditional desktop and mobile contexts the primary user interface is a screen, in IoT screens are rare or very small, which invalidate most of the traditional approaches. We examine how end-users interact with IoT products and how those products convey information back to the users, particularly \textit{`what is going on'} with regards to their data. We focus on understanding what the breadth of IoT, privacy, and ubiquitous computing literature tells us about how individuals with average technical expertise can be notified about the privacy-related information of the spaces they inhabit in an easily understandable way. In this survey, we present a review of the various methods available to notify the end-users while taking into consideration the factors that should be involved in the notification alerts within the physical domain. We identify five main factors: (1) data type, (2) data usage, (3) data storage, (4) data retention period, and (5) notification method. The survey also includes literature discussing individuals' reactions and their potentials to provide feedback about their privacy choices as a response to the received notification. The results of this survey highlight the most effective mechanisms for providing awareness of privacy and data-use-practices in the context of IoT in shared spaces.

\end{abstract}



\section{Introduction}

\begin{figure*}[t]
\centering

\includegraphics [scale=.7]{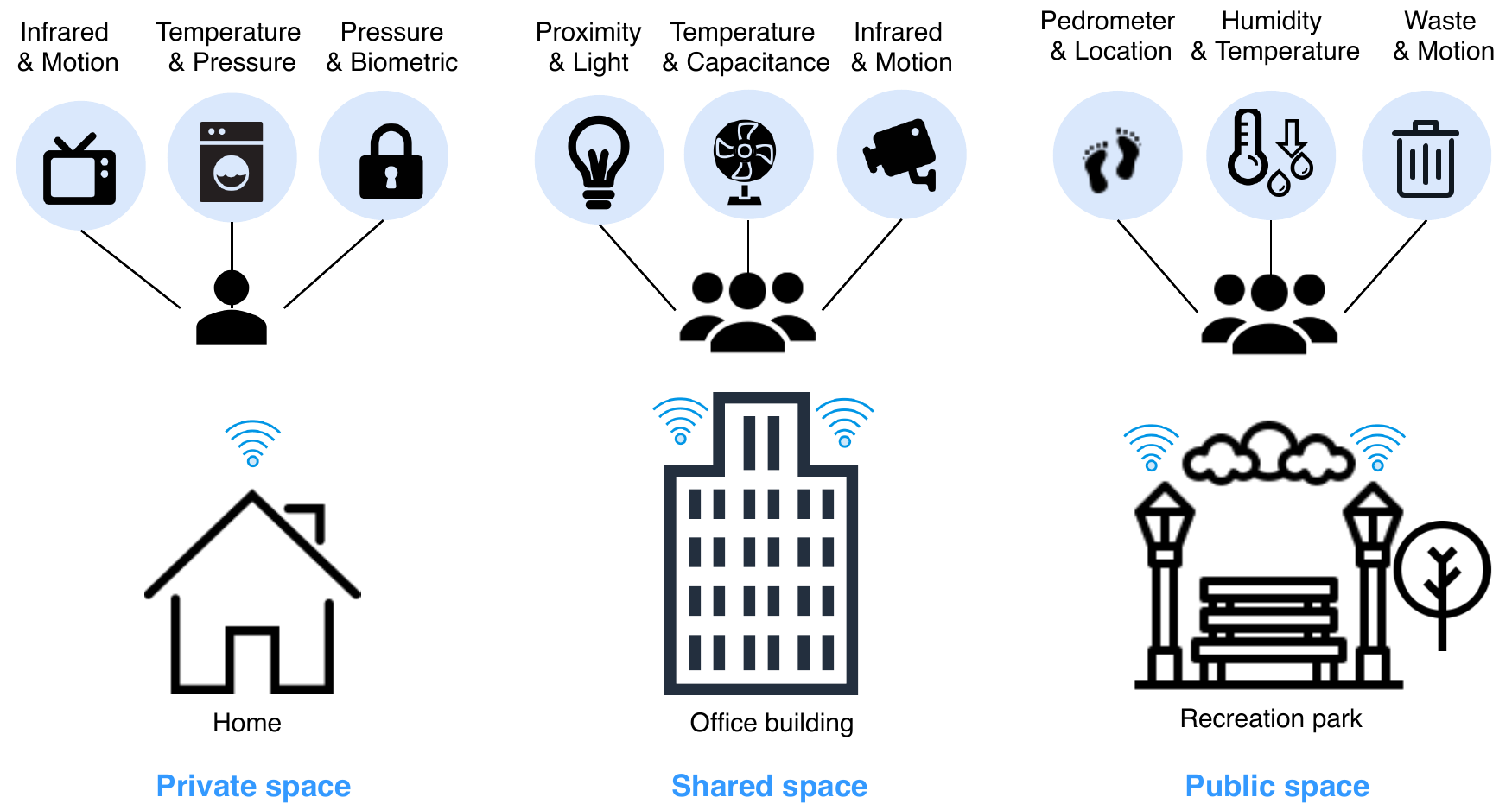}

\caption{IoT sensors employed in various spaces, which work on the collection of individuals data without their knowledge.} \label{fig13}

\end{figure*}

The built environment is currently undergoing a rapid transformation as shared spaces \cite{asquith2013understanding}\cite{edwards2005switching}, e.g., office building, transport, commercial and residential are being infused with sensors \cite{Stat2}, actuators, and interfaces and then labeled ``smart'' \cite{alam2012review}\cite{luria2017comparing}. An increasing number of people are interacting with data each day (estimated to be 5 billion in 2018 and growing to 6 billion by 2025 \cite{Stat3}). Further, fueled by the proliferation of IoT devices, it is estimated that in 2025 ``each connected person will have at least one data interaction every 18 seconds'' \cite{Stat3}. Each of these interactions has the potential to be recorded, analyzed, and shared. Today, the vast majority of those interactions are invisible. When a person walks into a ``smart'' shared space, they have no way of knowing what technology is in that space, what data it captures, and what happens to that data.  Figure \ref{fig13} depicts how different data can be collected about individuals in various spaces, without their knowledge.

 Spreading awareness poses a challenge in the IoT domain. That is due to three primary reasons: (1) the nature of the data the IoT device collects, (2) the nature of the people interacting with the IoT device, (3) and the nature of the service or task the IoT device provides/performs. 
 In some countries, shared spaces employing IoT technologies are obligated to provide privacy notices to notify people about the existing technology and its capabilities \cite{tzafestas2018ethics}\cite{weber2015internet}. However, most people lack awareness regarding the capture and use of their data \cite{zheng2018user}. Unfortunately, most privacy notices are represented in two forms: (1) a privacy notice sign, e.g., CCTV camera in operation, which might get unnoticed and reveals no information to the people about what happens to their data \cite{calo2011against}, (2) a privacy notice form, e.g., a document listing long privacy policies for the user to consent on, which is usually neglected \cite{acquisti2017nudges}\cite{gomulkiewicz1996brief}\cite{kaushik2018security}. Information disclosure is more complicated in the IoT domain due to their wide distribution and passive capability in collecting people information \cite{neisse2015agent}\cite{marannanstudy}. Consequently, organizations frequently complain about the complexity posed by the disclosure of their privacy policies \cite{acquisti2017nudges}.

 This survey discusses the available efforts in increasing users' awareness and applying them in the IoT domain. There are plenty of works that have been done in supporting this endeavour, such as \cite{leiser1989improving}\cite{haslgrubler2017getting}\cite{bodnar2004aroma}\cite{kaye2001symbolic}. However, based on our knowledge, most of them either address limited ways of enhancing user's awareness, require web-based tools, or are directed into specific, usually technical, users. Furthermore, there is a lack of having a basic development language that involves the necessary information an average user can understand. Consequently, there is also a lack of a formal interaction language between the user and the IoT sensor. Given that, in this survey, we did an intensive review of the available literature. We concluded the leading 5 factors needed to be considered while presenting any IoT sensor to the user, with an increased focus on the fifth factor, which is the notification method.

The importance of the fifth factor stems from the fact that we now live in a world of connected things, where notifications are everywhere \cite{pielot2014situ}. Smart devices and sensors now have the ability to generate and deliver numerous notifications in a matter of seconds \cite{atlam2020iot}. We passed the problem of not having a notification. The problem that arises now is how to maintain the privacy of the user using the notification method? Additionally, given the intensity of the notifications, the users receive every day, users' attention to the notifications tends to get affected, i.e., reduction in task execution \cite{b1998task}\cite{bailey2000measuring}\cite{horvitz2001notification}\cite{czerwinski2000instant}. Obviously, there is a lack of a unified model for notifying the users about their privacy.

 \subsection{End User Data Privacy Awareness}
\label{Individuals Data Privacy Awareness}
 
 In a shared space environment, individuals should be able to notice the IoT sensor with minimal to no efforts; they also need to be reminded that the IoT sensor still exists after noticing it. That is to say, if someone walks into the room and sees through a sign, for example, the CCTV recording camera, they probably would forget that something is recording them after 1 to 2 hours, and might perform a personal action that they do not wish to be recorded \cite{lau2018alexa}. One step forward in addressing this issue is the use of the Platform for Privacy Preferences Project (P3P) \cite{reagle1999platform}. P3P protocol was not designed for IoT, but its concept of using a simple way to notify the users about the visited website's privacy polices can be applied in the IoT domain.
 
 P3P allows users to specify their privacy preferences, giving them dominance on their data. By employing the P3P protocol, the users are aware that their data is being collected, used, stored, and retained. Similarly, in the IoT domain, when a proper notification method is applied, the users will be aware that their data is being collected, used, stored, and retained. These four factors, (i.e., data type, data usage, data storage, and data retention) have also been highlighted in other studies, such as \cite{shayegh2017toward}\cite{schaub2017designing} \cite{ naeini2017privacy}\cite{cha2018user} \cite{gluck2016short}\cite{langheinrich2002privacy} \cite{huang2020iot}\cite{jiang2004smart}. We build on that and discuss each of theses factors with an emphasis on the fifth factor in which we surveyed the available notification methods. P3P protocol sets the stone in involving the users into preserving their data online, and it gives them choices on where and with whom they could share their data. Having a privacy awareness model using a predefined logic enhances the ability of individuals' notification criteria \cite{amaratunga2002quantitative}.
 
 One of the primary purposes of an awareness model is to notify people about the existence of an IoT sensor in their vicinity, which could be done in various ways. P3P has the warning technique that notifies users about a conflict between their privacy policies and the policy of the websites they are trying to reach \cite{p3p1}. Furthermore, P3P grants the users a choice of either rejecting visiting the website or proceeding despite the conflict \cite{p3p1}. To have aware IoT users, they need to be continuously warned about the IoT sensors in operation \cite{lau2018alexa}. Adding to that, the users should also know the type of data the IoT sensor is collecting, where the data is stored, in what the data is being used, and for how long it is retained \cite{naeini2017privacy}. These four factors combined with a simple notification method can build a privacy-aware IoT notices in shared spaces. 
 
 Although the P3P protocol failed \cite{p3p1}\cite{p3p2}, it's concept in  helping the users to get more control on their data is crucial in the IoT domain. In Table \ref{table:1}, we present a comparison between the IoT domain (current and proposed) and the P3P protocol. Since this survey is mainly about awareness, we focus on discussing (1) Delivery language: represented in the paper in the notification methods section, (2) Data collection: represented in the paper in the data type section, (3) Data usage: represented in the paper in the data usage, storage and retention sections.

\begin{table*}[t!]

\small

\begin{center}

 \caption{Differences between the IoT domain (current and proposed) and the P3P protocol regarding the user data control.}
 \label{table:1}
 
\begin{tabular}{ p{2.6cm} p{3cm} p{3.5cm} p{3.5cm} } 

 \hline 

& \centering P3P &{IoT: Sensors (now)} & IoT: Sensors (Proposed)
\\ 
\hline \hline

 Choice & User can define his privacy policy & User is required to agree to the privacy policy for the chosen IoT device & User can modify the privacy policy according to his needs  \\ 
& Difficult for average user &Average user can set it  & Average user can set it \\
  \hline
 Setup language & Website: XML User: user agent  & N/A Required to agree cannot setup & Short instructions e.g., buttons
  \\ 
& Difficult for average user & & Average user can set it \\
 
 \hline
   Delivery language & Lengthily terms \& polices & Lengthily terms \& polices & Precise terms, short notifications   \\ 
& Difficult for average user &Difficult for average user  &Average user can understand \\

 \hline
 
 Data Collection & Web data collected by the website Cookies,
user provided data: emails, birth date etc.
 & Personal data either provided by the user, collected by the device or both & Personal data either provided by the user, collected by the device or both \\
 

& User usually is not aware of the collection &User usually is aware of the collection, but does not have control & User usually is aware of the collection and have control\\
 
 \hline
 Data usage & Users web data is being used in other services. e.g., improve browsing habits, statistics, ads, etc.  & Users data is being used or sold to other services. e.g., improve sensors, statistics, ads, etc. & Users data is being used or sold to other services. e.g., improve sensors, statistics, ads, etc. \\

& User usually is not aware &User usually is not aware, usually have control & User usually is aware and have control\\
 
 \hline
 People using the service & Adults \newline(usually people with computer background) & Varying users \newline(including elderly, children) & Varying users \newline(including elderly, children) \\ 
 \hline

\end{tabular}

\end{center}

\end{table*}

 \textbf{Existing research:} Individuals (End Users) privacy awareness is a deep-rooted topic that has captured the researchers' interest from a long ago. There have been plethora pieces of research which investigated the importance of individuals' privacy awareness in shared spaces. The difficulty faced by individuals when making privacy decisions and the hurdles faced by developers trying to comply with privacy policies is discussed in \cite{lodge2019privacy}\cite{vallejo2018kids} \cite{seymour2020informing}\cite{rafferty2017towards}. Individuals behaviour and how individuals' privacy awareness in the context of IoT can be elevated has been studied and analysed by \cite{lee2019confident}\cite{zheng2018user}\cite{hwang2018respectful}\cite{niemantsverdriet2019designing}.  Also, studies that support individuals privacy awareness through studying users preferences and reactions to notifications are presented in \cite{acquisti2017nudges}\cite{mehrotra2017intelligent}\cite{orth2019designing}\cite{chhetri2019towards}\cite{rozendaal2019objects}\cite{perera2013context}. Moreover, the use of natural interfaces has shown its effectiveness in attracting individuals attention \cite{nass2001does}\cite{shin2013defining}, which can be adopted in the IoT domain. As defined in \cite{meulen2017gartner}, the IoT is a network of physical objects, in  which  each  object has the ability to sense, similar to human sensors (we elaborate on this more in Section \ref{Data Type}).
 
 \textbf{Novelty of this Paper:} It is important to note that none of the existing surveys has reviewed the relation of IoT sensors to human sensors. The novelty of this survey is presented in categorising the IoT sensors based on human sensors, allowing simplifying and understanding the way an IoT device is collecting and using individuals’ data \ref{Data Type}. In addition, this paper discusses the essential awareness factors with a focus on the notification methods, in which we classified a considerable number, 31 studies, of prior works into four human-related classifications \ref{Notification Methods}. Our objective is to identify major factors that the IoT domain needs to support especially towards creating a privacy-aware environment. For this end, after conducting a thorough literature review, we developed a taxonomy of common factors and present multiple use case scenarios. We compared several research findings and efforts as well as identifying research trends and gaps and highlighting research challenges.

    The General Data Protection Regulation (GDPR) act \cite{R1} and the California Consumer Privacy Act (CCPA) \cite{ccpa2020california}, have confirmed the importance of people awareness regarding the use of their personal data. However, until now, and based on our knowledge, there is no known technology or used technique that notifies people about the use of their data in a short and direct form. For that this paper contributes to the following:

\begin{itemize}
\item Assess the available techniques, protocols, models, and literature pertaining to individual's data privacy awareness. 
\item Propose and use a taxonomy to categorise available models, as well as to compare and contrast past approaches.
\item Review data privacy factors collected by most IoT devices.
\end{itemize}

\textit
Paper structure: The paper is divided into five sections and is structured as follows: The used and followed methodology is presented in Section \ref{methodology}, which includes the data extraction method and a list of the used search queries. The main content of the survey is presented in Section \ref{Privacy Managed Infrastructure}. It is divided into five main subsections, where each discusses one of the main factors pertaining to individuals' awareness. The five subsections are Data type, data usage, data storage, data retention, and notification methods. In Section \ref{Discussion}, we include a discussion about the IoT user awareness and presents the main available gaps in this area. Section \ref{Research Challenges and Opportunities} discuses the research challenges and opportunities. Lastly, Section \ref{Conclusion} concludes the privacy awareness survey.

\subsection{Methodology}
\label{methodology}
 This survey is a result of a thorough review of the literature in the area of awareness. In this paper, we draw from the results and findings in this area to deliver an organised summary of the available notification methods that are (or can be) incorporated into the IoT domain. Our research is inspired by a research challenge on the different kinds of notification methods which was discussed in \cite{hong2017privacy}. The authors in \cite{hong2017privacy} mentioned a project called “signifiers” \cite{springrsensor}\cite{summersensor} that explored two notification methods, namely “visual and audio”, from which we built our initial search queries. To build this survey, we followed the PRISMA methodology \cite{liberati2009prisma} (as shown in Figure \ref{fig:15}), and performed several steps as follows:

  \begin{figure}[t]
\centering
\includegraphics [ scale=.4]{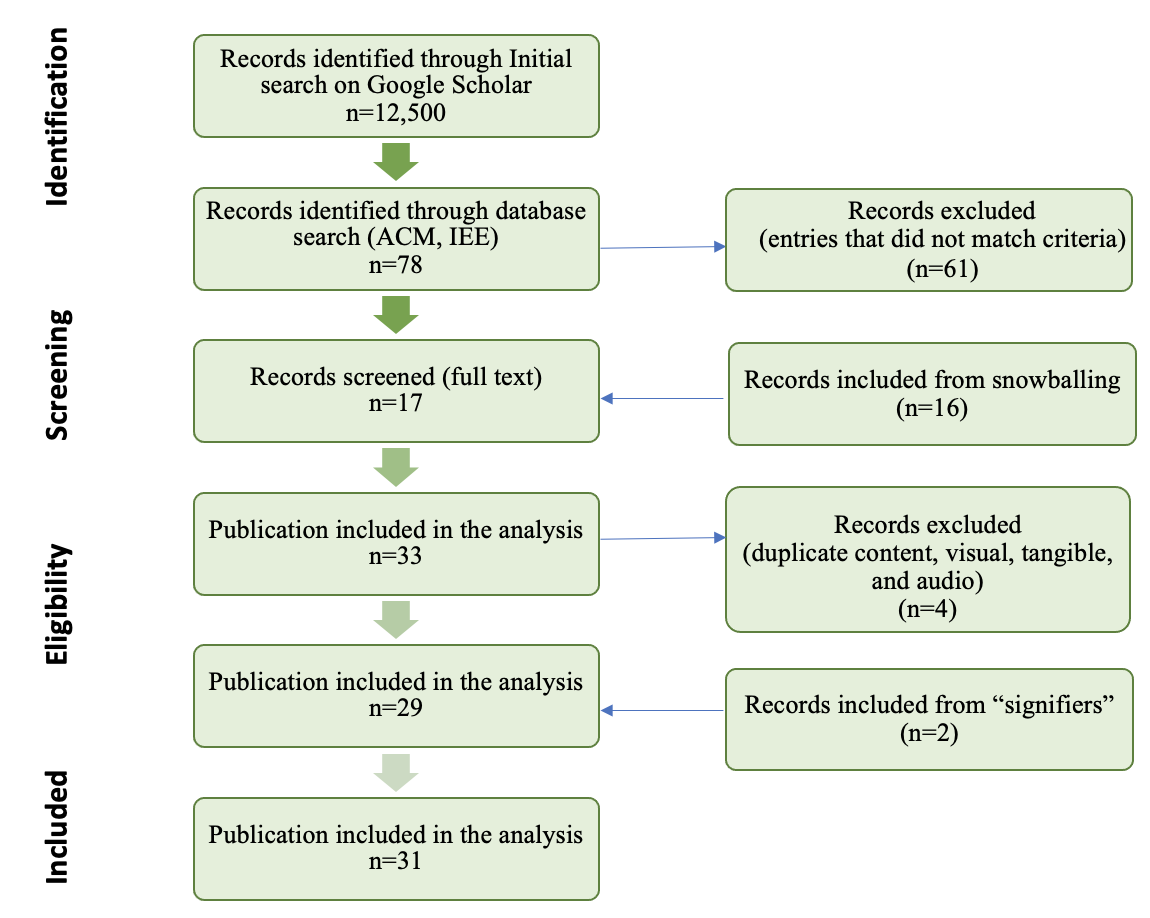}

\caption{Flowchart of the publications section process adapted from PRISMA \cite{liberati2009prisma}} \label{fig:15}

\end{figure}

First, we conducted an exploratory automatic search on Google scholar to avoid bias of any specific publisher \cite{wohlin2014guidelines}, and to make it possible to identify the relevant databases with more scientific support. Our initial search queries included the keywords ("visual" OR "light" OR "warn" OR "tangible" OR "wearable" OR "audio" OR "sound" OR "olfactory" OR "vibration") filtered by ((“notification” AND “aware”) AND ("IoT" OR “internet of things”)). From 2011 to 2019, 12,500 articles were retrieved. Most of the retrieved articles were either from ACM Digital Library or IEEE Xplore Digital Library.

Next, we performed a manual search on both ACM Digital Library and IEEE Xplore Digital Library using more specific keywords as tabulated in Table \ref{table:5}. For the online libraries, the studied time period was the same as Google Scholar from 2011–2019, in which 78 articles were retrieved (48 results from ACM, 30 results from IEEE).

We then screened the articles manually following an inclusion criteria that mandated they (1) mentioned at least one of the data privacy factors (i.e., data type, data purpose, data storage, data retention), (2) involved strategies for human interaction and awareness, (3) described actual design or results, (4) can serve IoT domain. 17 articles satisfied the criteria. 

Lastly, we performed snowballing (backward) on the papers from manual search, and included 16 more articles. Further, we excluded 4 articles which contain duplicated content. In total, we included 31 articles in our survey (13 from manual search, 16 from snowballing, and the 2 "signifiers" articles \cite{springrsensor} \cite{summersensor}. We furthure e The literature referenced in this survey span a wide period (1997 to 2019), with a focus on the papers published in the last 10 years. Doing so was to ensure the broad coverage of the available literature while maintaining an up to date research.

\begin{table*}[h]
\small
\begin{center}
\caption{Search queries and terms used in acquiring the literature either from Google Scholar or online libraries.}
\label{table:5}

\begin{tabular}{p{.15\textwidth}||p{.8\textwidth}}
\hline
Category                & Search queries and terms                                                                                                                                                                                                                                                                                                                    \\ \hline
\centering General search queries  & \begin{tabular}[c]{@{}l@{}}
\texttt {"notification"}\\ 
\texttt {"aware"}\\  
\texttt {"IoT" OR "internet of things" }\\ 
\texttt{"data type" OR "data purpose" AND (combination of the above)} \\ 
\texttt{"data storage" OR "data retention" AND (combination of the above)} \\ 
\texttt {"shared space" AND (combination of the above)}\\
\texttt{"visual" OR "light" OR "warn" AND (combination of the above)} \\ 
\texttt{"tangible" OR "wearable" AND (combination of the above)} \\
\texttt{"audio" OR "sound" AND (combination of the above)} \\ 
\texttt{"olfactory" OR "vibration" AND (combination of the above)}
\end{tabular}                                             \\ \hline
\centering Specific search queries & \begin{tabular}[c]{@{}l@{}}In the three used databases: Set the language to English \\ Google Scholar: all-initial:  \texttt{(("visual" OR "light" OR "warn" OR "tangible" }\\
\texttt{OR "wearable" OR "audio" OR "sound" OR "olfactory" OR "vibration")} \\
\texttt{AND ((“notification” AND “aware”) AND ("IoT" OR “internet of things”))} \\ IEEE: 
\texttt{("(All Metadata":notification) AND ("All Metadata":aware)} \\
\texttt{AND ("All Metadata":internet of things OR IoT)} \\ ACM:  \texttt{AllField:(notification) AND Abstract:(internet of things) AND }\\
\texttt{Abstract:(aware) AND AllField:(visual) OR AllField:(light) OR }\\
\texttt{AllField:(warn) OR AllField:(tangible) OR AllField:(wearable) OR}\\
\texttt{AllField:(audio) OR AllField:(sound)OR AllField:(olfactory) OR}\\
\texttt{AllField:(vibration)}\\ \end{tabular} \\ \hline
\end{tabular}
\end{center}

\end{table*}

\section{Privacy Managed Infrastructure}
\label{Privacy Managed Infrastructure}

The widespread of IoT sensors makes them an essential part to go with everyday life activities \cite{asquith2013understanding}. Users of all ages have an interaction with at least one IoT sensor daily, which in return relays on collecting and processing their information \cite{Stat3}. It is an alarming issue of how could these sensors collect and process a massive number of users' data without any consciousness from the users \cite{atlam2020iot}. Users might not have a clue about the IoT sensor in the room and may not know its data collection and processing capabilities \cite{zheng2018user}. 

This section provides an assessment of the literature that discusses the available or proposed mechanisms that looked into increasing the users' awareness regarding IoT exposure. By the term IoT exposure here we mean: the existence of one or more IoT sensor, the sensor is collecting the user's information, processing this information (retaining them, sending them somewhere, selling them to other parties, etc.) without the user notice.

\begin{figure}[t]
\centering
\includegraphics [ scale=.5]{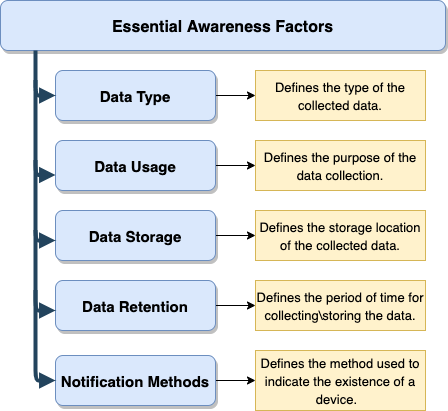}

\caption{Essential awareness factors that must be incorporated in a privacy notice for the user to be aware of the data collection and processing done through an IoT sensor.} \label{fig:9}

\end{figure}

As mentioned in section \ref{Individuals Data Privacy Awareness}, to increase the individuals' awareness regarding IoT privacy, a privacy notice needs to satisfy five main factors, i.e., data type, data usage, data storage, data retention, and notification methods. Figure \ref{fig:9} presents a definition of each factor. We defined these factors based on prior works that study the individuals' privacy \cite{naeini2017privacy}\cite{barua2013viewing} \cite{klasnja2009exploring}\cite{lederer2003wants}\cite{lee2016understanding}\cite{lee2017privacy}\cite{leon2013matters}. First, the user needs to be aware of the collected data type, such as audio, video, and(or) temperature data. Second is data usage, which defines the purpose of the data collection, whether for telemarketing, energy-saving, entertainment, security, improving health, or other reasons. Following that is data storage. Can the collected data be stored within the device, or is it possible to store them company-wide or even in third-party storage? The fourth factor that requires the user's attention is data retention, which lays out the data collection time and frequency. Lastly and importantly, an effective notification method must be applied for the user to easily notice an IoT sensor in the vicinity.

\subsection{Data Type}
\label{Data Type}
\begin{figure*}
	\centering
	\vspace{-24pt}
	\includegraphics [scale=.6]{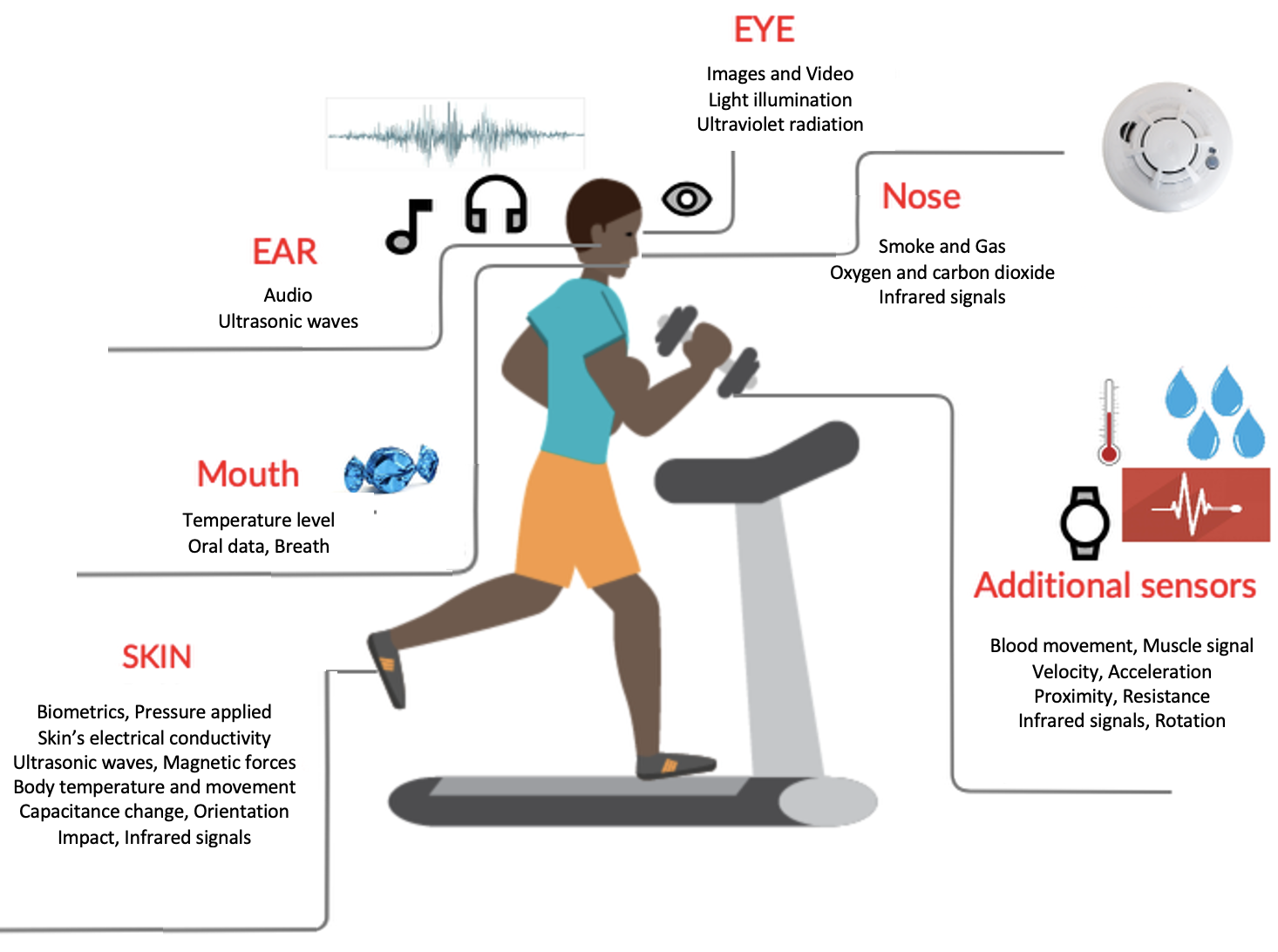}
		\vspace{-14pt}
	\caption{IoT data types categorised using human sensors, each human's sensor represents a sample of the data collected by an IoT sensor.}\label{fig2}
	
\end{figure*}

To begin, we will start analysing the pieces of literature of the first factor of the user's awareness. In the context of IoT, there exists a multiplicity of sensors, which each collects one or more data types. Collected data will then be processed to deliver a variety of services to the user. The provision of these services depends mainly on analysing the data that are collected from human activities, e.g., body posture and movement. Although the possession of IoT sensors gives the user a luxury feeling along with higher productivity and automation \cite{stojkoska2017review}, users usually are not aware of the type of data collected by these devices \cite{zheng2018user}. Moreover, users do not know what other information could be derived from the collected data to provide them with the desired level of service they acquire. What is worse is that some of the collected data are not required at all to deliver a service to the user, giving the sense that it is being collected for other purposes. In \cite{hong2017privacy}, for instance, they have stated that around half of smartphone apps are collecting location data without the need to; They are, in fact, collecting it for the use of a third-party library. In this section of the survey, we are outlining the major data types that have been discussed in previous literature and that relate to a shared space environment.

In a shared space environment, the types of sensors that are used are directly or indirectly related to enhancing the efficiency of the user's daily activities. These sensors, despite their heterogeneous, share the collection of common data types to provide users with the task they require. Given that, it can be seen that the IoT sensors, when working together, can mimic the human's sensing ability in collecting the information \cite{someya2013building} \cite{karimi2015role}. Through the sensors in the human body, the brain receives information and make decisions. A typical scenario is, a person sensing a burning smell from the kitchen, will immediately turn off the stove. The human sensor here is the nose, the action is turning off the stove, and the collected data type is the smell. In the IoT context, sensors have a similar working schema. A smoke detector being the sensor, for instance, will trigger an alarm and cause the stove to be turned off (action) when smoke is detected (collected data). 

Gartner has defined the IoT as a network of physical objects, i.e., IoT sensors, in which each object has an embedded technology that senses and(or) interacts internally and(or) externally \cite{meulen2017gartner}. This is similar to the human body, in which each of the human five sensors interacts internally or externally with the brain to perform an action based on sensation \cite{meulen2017gartner}. Consequently, the different IoT sensors through the data they collect can sense and perform an action (i.e., the smoke detector example in the previous paragraph). The similarity in sensation maps the IoT sensors to human sensors. Doing so offers a more natural way of relating the IoT data to its sensor. Using natural interfaces has shown its effectiveness in attracting users' attention \cite{nass2001does}\cite{shin2013defining}. In Figure \ref{fig2}, we show link the common data types to their five primary human sensors, namely: eyes, nose, ears, skin, and mouth. In addition, there are many indirect human sensors, such as blood vessels, which sense the amount of blood to get medical diagnoses. Table \ref{table:2} further tabulated the information presented in Figure \ref{fig2} adding to them two additional rows, i.e., the indirect sensors and and the external sensors. With referring to Gartner definition mentioned earlier in this paragraph, it is worth noting that, in a shared space, beside sensing the users' activities, there are other sensors that senses externally, e.g., sensing the environment, not the user, to provide the user with a better experience. An example of these sensors is the smart thermostat, which can detect the room temperature and turn on the heating system in the house accordingly. Likewise, the human brain uses more than one sensation to make decisions; many IoT devices might also collect data using more than one sensor to provide accurate information. As seen in Table \ref{table:2}, we present some examples of the sensor types and their IoT application.

\begin {table*}
   \small
 \caption{Data types collected through IoT sensors and their equivalent human sensor based on \cite{apthorpe2017smart}\cite{Seedstudio}\cite{bai2018magnetoresistive}\cite{sensor2}.} 
 
\label {table:2} 
\begin{tabular}[t]{p{.12\textwidth} p{.25\textwidth} p{.28\textwidth} p{.3\textwidth}}

 \hline \hline

& Sensor Type & Data Type Detected & IoT Device Application
\\ \hline \hline

\center{Ear} & \vspace{-3mm} 
\begin{itemize} [leftmargin=.25cm]
\item Sound Sensor
\vspace*{-\baselineskip}
\end{itemize} 

 & \vspace{-3.5mm}  \begin{itemize} [leftmargin=.25cm]
     \item Audio
     \item Ultrasonic waves
     \vspace*{-\baselineskip}
 \end{itemize} 
  &  
 \vspace{-3mm} \begin{itemize} [leftmargin=.25cm]
     \item Voice recognition systems
     \item Distance measurements
     \vspace*{-\baselineskip}
 \end{itemize} \\ 

\hline

\center{Eye}& 
 \vspace{-3mm} 
 \begin{itemize} [leftmargin=.25cm]

 \item Camera Sensor
 \item Colour Sensor
 \item Light Sensor
 \item Fire Sensor
 \vspace*{-\baselineskip}
 \end{itemize}
 & \vspace{-3mm} 
  \begin{itemize} [leftmargin=.25cm]
     \item Images and Video 
     \item Lights illumination \newline (Colours Photodiodes)
     \item Ultraviolet radiation
     \vspace*{-\baselineskip}
 \end{itemize} 	 &
  \vspace{-3mm} 
   \begin{itemize} [leftmargin=.25cm]
     \item Monitoring systems
     \item Face recognition systems
     \item Smart lightning systems
     \vspace*{-\baselineskip}
 \end{itemize} \\ 

  \hline
  
\center{Nose} & 
 \vspace{-3mm} 
\begin{itemize} [leftmargin=.25cm]
\item Smoke Sensor
 \item Gas Sensor
 \item Odour Sensors
 \vspace*{-\baselineskip}
\end{itemize}
 &  \vspace{-3mm} 
 \begin{itemize} [leftmargin=.25cm]
     \item Smoke and Gas 
     \item Oxygen and carbon dioxide levels
     \item Infrared signals
     \vspace*{-\baselineskip}
 \end{itemize} 	 &  \vspace{-3mm} 
 \begin{itemize} [leftmargin=.25cm]
     \item Air quality monitoring
     \item Smoke detection systems
     \item Smart Gardening
     \vspace*{-\baselineskip}
 \end{itemize} \\ 

\hline
  
\center{Mouth} & 

 \vspace{-3mm} \begin{itemize} [leftmargin=.25cm]
\item Level and Temperature Sensor
\item Alcohol Sensor
\item Moisture Sensor
\vspace*{-\baselineskip}
\end{itemize}

 &  \vspace{-3mm} \begin{itemize} [leftmargin=.25cm]
     \item Temperature level
     \item Oral data
     \item Breath
     \vspace*{-\baselineskip}
 \end{itemize} 	 & \vspace{-3mm}  \begin{itemize} [leftmargin=.25cm]
     \item Alcohol monitoring systems
     \item Diet monitoring systems
     \item Food tasting systems
     \vspace*{-\baselineskip}
 \end{itemize} \\

\hline
  
\center{Skin} &  \vspace{-3mm} 
\begin{itemize} [leftmargin=.25cm]
\item Touch Sensor

\begin{itemize}
    \item Force Sensor
\end{itemize}
\item Skin Sensor \item Electromyography
\item Proximity sensor
\item Temperature Sensor
\item Vibration Sensor
\item Line Finder
\item Distance sensor
\vspace*{-\baselineskip}
 \end{itemize}
 
 &  \vspace{-3mm} 
 \begin{itemize} [leftmargin=.25cm]
     \item Biometrics
     \item Pressure applied 
     \item Skin's electrical conductivity
     \item Ultrasonic waves
     \item Magnetic forces
     \item Body temperature
     \item Body movement
     \item Capacitance change 
\item Infrared signals
\item Orientation
\item Impact
\vspace*{-\baselineskip}
 \end{itemize} 	 &  \vspace{-3mm} \begin{itemize} [leftmargin=.25cm]
    \item Fingerprint scanner
\item Galvanic skin response
\item Medical systems
\item Security systems
\item Smart toys
\item Automatic Lightning 
\item Smart appliances
\item Vehicles seat monitors
\item Smart vacuum
\item Activity trackers
\item	Smart transportation
\item Smart locks
\vspace*{-\baselineskip}
 \end{itemize} \\ 

\hline

\center{Additional sensors e.g., blood vessels} & 
 \vspace{-3mm} 
\begin{itemize} [leftmargin=.25cm]
\item Heart rate sensor
\item Optical Sensors
\item Gesture Sensor
\item Rotary Sensor
\item Motion Sensor

\begin{itemize} [leftmargin=.25cm]
\item Gyroscope
    \item Accelerometer
    \item Magnetometer
\end{itemize}
\vspace*{-\baselineskip}
 \end{itemize}

 & \vspace{-3mm}  \begin{itemize} [leftmargin=.25cm]
 
\item	Blood movement
\item	Muscles Signal
\item	Velocity (Speed)
\item	Acceleration
\item	Proximity
\item	Resistance
\item	Infrared Signals
\item	Rotation (direction)

\vspace*{-\baselineskip}
   
 \end{itemize} 	 &  \vspace{-3mm} \begin{itemize} [leftmargin=.25cm]

 \item	Sleep monitors
\item	Heart-rate monitors
\item	Wearable sensors
\item	Baby monitors
\item	Blood sugar monitor
\item	Transponders on animal
\item	DNA analysis devices
\item	Smart navigation systems
\vspace*{-\baselineskip}

 \end{itemize} \\ 

\hline

\center{Sensors not directly related to human sensors} & 
 \vspace{-3mm} 
\begin{itemize} [leftmargin=.25 cm]
\item Temperature Sensor
\item Humidity Sensor
\item Water Sensor
\item Turbidity Sensor
\item Ultraviolet Sensor
\item Dust sensor
\vspace*{-\baselineskip}
 \end{itemize}

 &  \vspace{-3mm}  \begin{itemize}[leftmargin=.25cm]
 
\item	Temperature and Humidity level
\item	Atmosphere pressure
\item	Capacitance change 
\item	Ultraviolet radiation
\item	Light
\item	Slop
\item	Dissolved solids
\item	Hydrogen ion 
\item	Dust concentration
\vspace*{-\baselineskip}

 \end{itemize} 	 &  \vspace{-3mm} \begin{itemize}[leftmargin=.25cm]
\item	Tank systems 
\item	Smart appliances
\item	Sewage systems
\item	Liquid sensing applications 
\item	Pharmaceuticals 
\item	Dyeing process
\item	Elevators systems
\item	GPS
\item	Smart meter
\item	Smart thermostat
\vspace*{-\baselineskip}
 \end{itemize} \\ 
 
\hline

\end{tabular}
\end{table*}


\subsection{Data Usage}

The pervasive nature of IoT devices created new ways to use the data. As a result, these new ways fetched new privacy challenges, in which it becomes essential for the user to know the purpose of the IoT data collection. Average users, despite their possession of one or more IoT devices, think that the devices are using their data only to provide them with a better experience. User’s benefit is undoubtedly one of the primary purposes of the usage of the sensed data. However, there is a good deal of other data usages purposes that neither the user nor the developer – in some scenario- are aware of \cite{shen2019iot}\cite{balebako2014privacy}. To make matters worse, besides the basic usage of the collected data, additional information could be inferred from the collected data and be used to build more knowledge \cite{ren2019information}. In the next two subsections, we conclude the main purposes of data usages that have been brought up by prior works. 

\subsubsection{The primary purpose of data usage: }
\begin{figure*}[t!]
\centering
	\includegraphics [scale=.6]{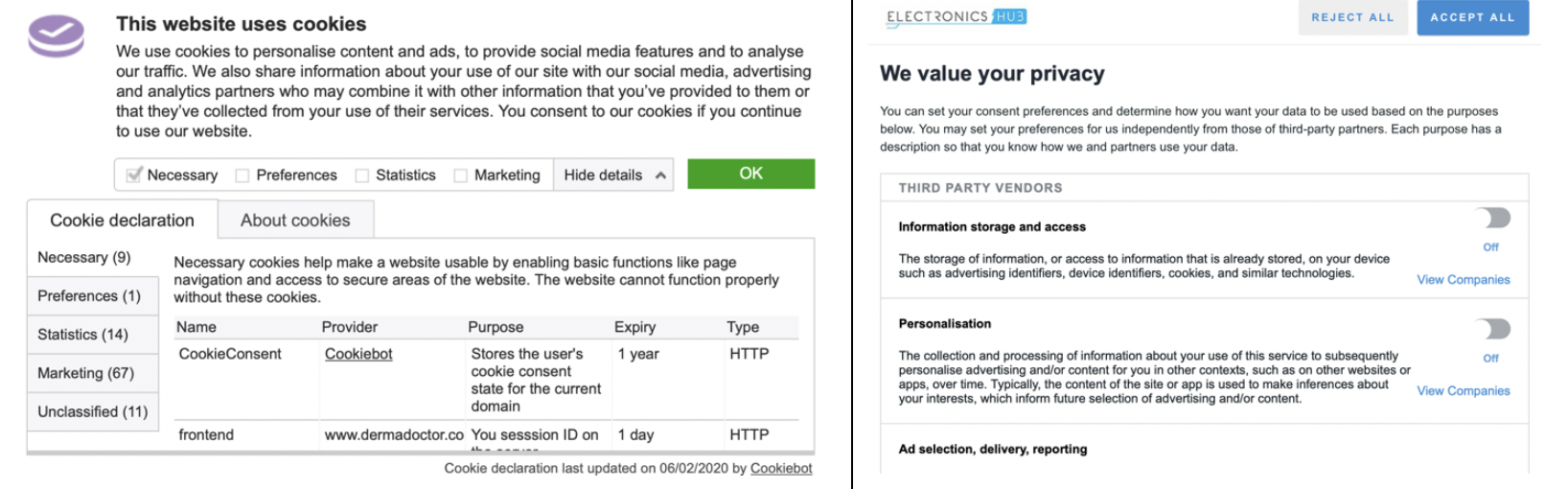}
	\caption{Sample of two websites' privacy policy purpose, the user must agree in order to browse the chosen website \cite{Dermadoctor}\cite{HowStuffWorks}.}

	\label{fig4:mesh4}
\end{figure*}

When it comes to defining the purpose of collecting the data, there is a considerable variation. The purpose of using the collected data would mainly depend on the device that is collecting it. Usually, when the user purchases a particular sensor, the privacy policy attached to the device would mention one or more specific purposes of the data usage. That is similar to the privacy policy on the web, which users habitually ignore and accept without any further reading. Figure \ref{fig4:mesh4} is a sample of the purposes specified by one website. 

We have concluded the main purposes of shared spaces data usages in Table \ref{table:3}. We selected one IoT device from each section of Table \ref{table:2} based on the common data usage purposes specified in previous literature \cite{hong2017privacy}\cite{naeini2017privacy}\cite{acquisti2017nudges}\cite{cila2017products}\cite{cardboard}. It is worth noting here that the purposes classified in Table \ref{table:3} are the abstract purposes defined by the device, which usually do not reveal information of what is being done with the data. For the device to comply with its standard privacy policy, it must specify the data collection purpose. Manufactures usually tend to use a dim view when presenting the purpose of the data collection \cite{chu2018security}. For example, manufacture producing a smart smoke detector might specify in the privacy policy that they are using the user data to improve research and analytic, which will help in providing better user experience. However, the underlying mechanism is different. The manufactures are collecting the users activities, such as how often and for how long did they smoke, how many people are smoking, did the smoke comes from a cigarette or from another burning object, etc. Such purposes of the data collection are usually not included in the privacy policy \cite{chu2018security}. This information is usually referred to as the inferred knowledge of the data collection purpose, which is further described in the next subsection.

\begin{table*}[t!]
\begin{center}

\caption{Sample of data usage purposes for: User, IoT device, and Manufacture respectively.}
\label{table:3}
\begin{tabular}{ p{4 cm} || p{3.5cm} || p{3.5cm}} 

 \hline 

User Purpose & IoT Device Application & Manufacture Purpose\\ 
\hline \hline

Improve safety
\newline Improve security 
\newline Improve health 
\newline Energy saving 
\newline Improve spending 
\newline Entertainment
\newline Improve lifestyle experience

&

Voice recognition systems
\newline Smart lightning systems
\newline Smoke detection system
\newline Alcohol monitoring systems
\newline Security systems
\newline Heart-rate monitors
\newline Smart thermostat

&
Improve advertisements
\begin{itemize}
    \item Targeted ads
\end{itemize}

 Improve productivity 
\newline Increase revenue 
\begin{itemize}
    \item Improve selling
    \item Improve spending 
    \item Reselling
\end{itemize}

 Improve research 
\newline Improve analytic 
 \newline Improve statistics
\newline Improve security
\newline Improve safety 
\newline Improving health care
 Surveillance \\

 \hline
\end{tabular}
\end{center}

\end{table*}

\subsubsection{The secondary purpose of data usage:}
\begin{figure*}[t]
\centering
\includegraphics [scale=.4]{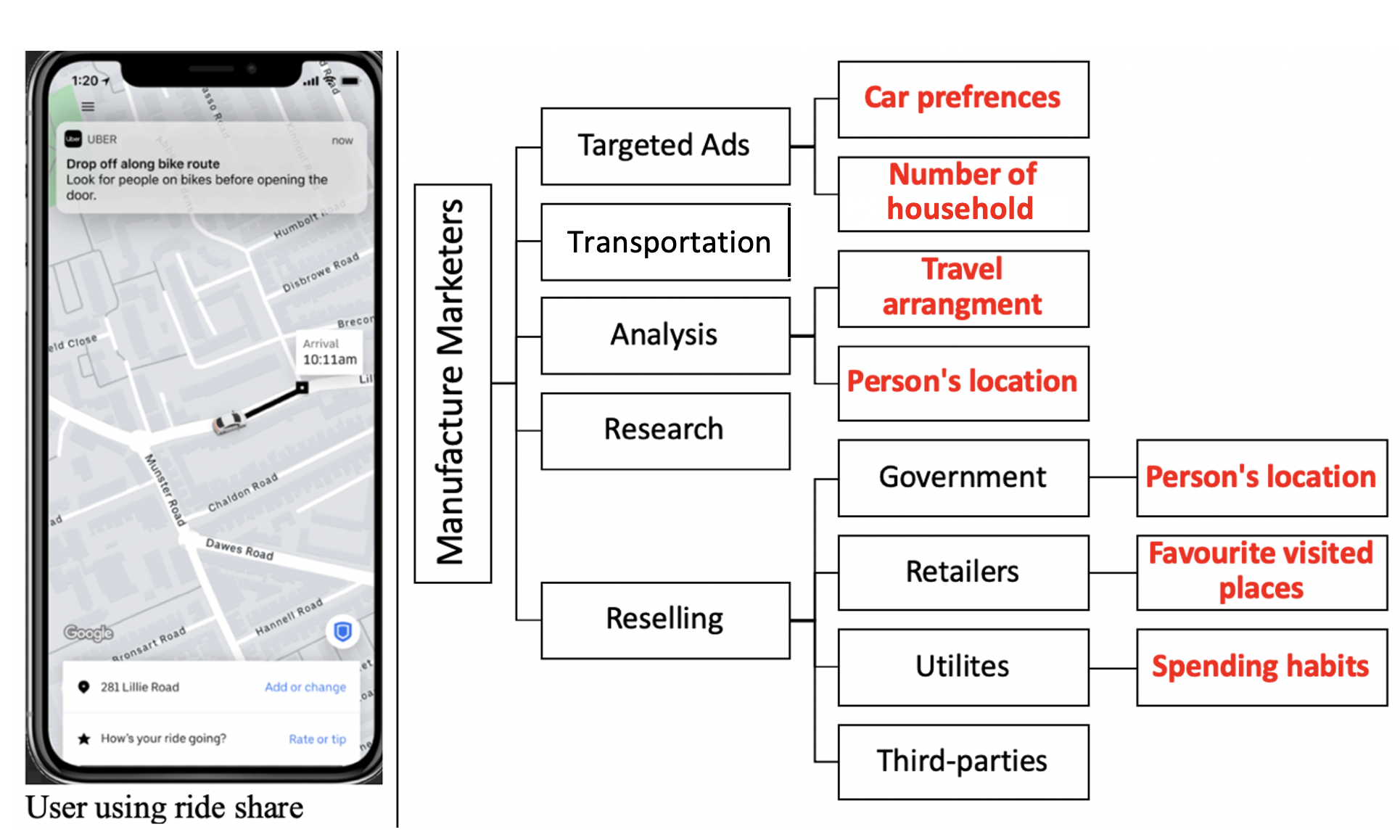}
\caption{Sample of a ride share app, showing how an additional knowledge can be inferred from the specified data usage purposes. The red boxes represents the inferred knowledge, and the black boxes represent the specified data usage purposes. }

 \label{fig5:mesh5}
\end{figure*}

 As described above, the purpose of using the collected data does not only cover the abstract meaning of improving research, for instance. It, however, spans a much wider area. The more data the device collects, the more knowledge it will have and can build, where the accumulation of the knowledge could lead to building a complete human profile. Figure \ref{fig5:mesh5} depicts how an inferred knowledge can be determined from a simple ride share application. To describe the value of the collected data, let us here provide two scenarios to show how the inferred knowledge could benefit the service providers and affect the individual's data privacy.

\begin{figure*}[t]
\centering
\includegraphics [ scale=.4]{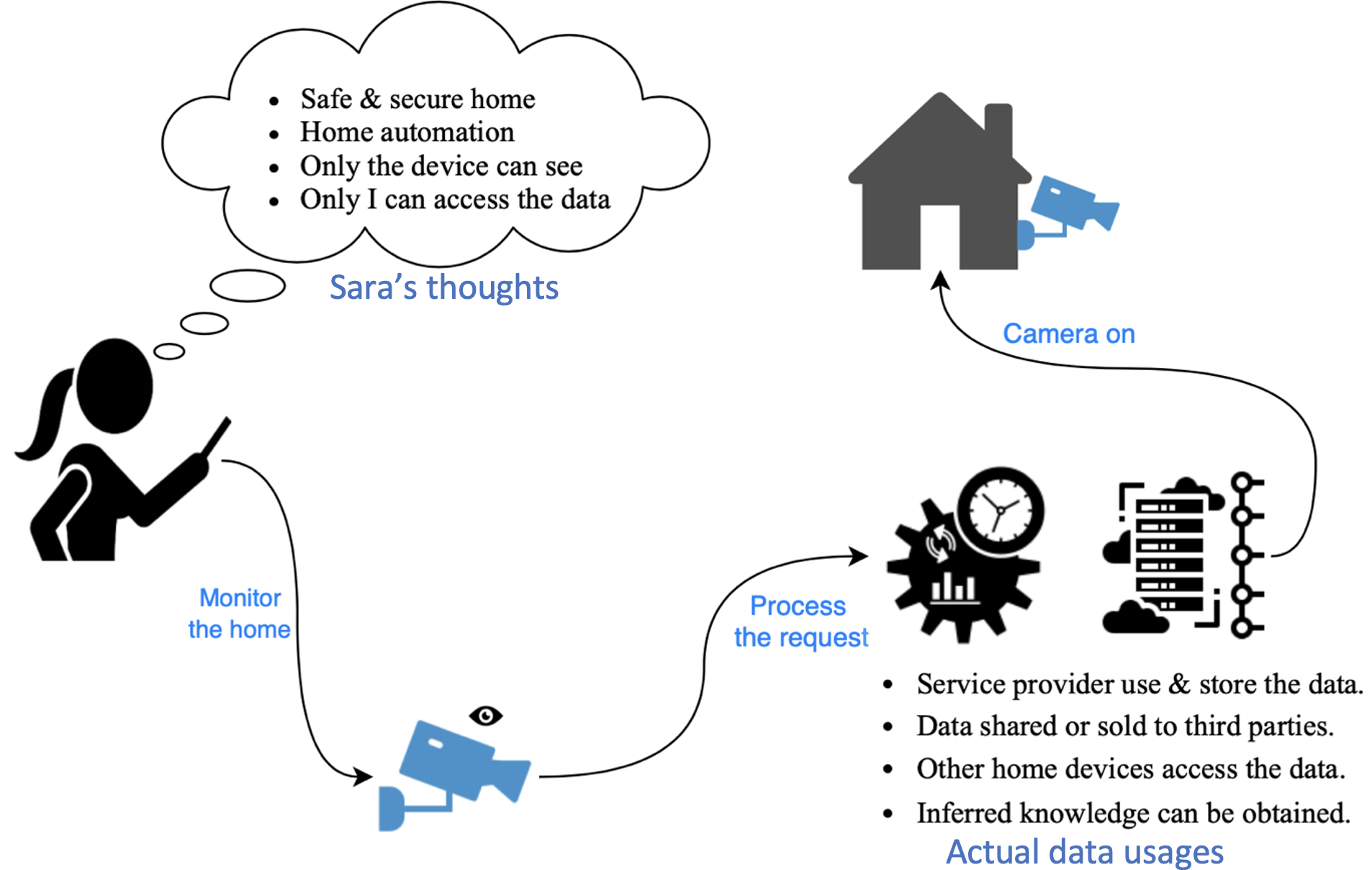}

\caption{Data usage use case scenario of a security alarm system including the user expected data processing and the actual data processing.} \label{fig:12} 

\end{figure*}

1) Security alarm systems: Sara is a frequent traveller, and therefore needed to monitor her home instantly. She is using a monitoring camera that can take images and stream videos whenever it detects motion. The recordings of her camera travel through different nodes until it reaches the application that provides her with the remote monitoring feature. Theses nodes include but not limited to third party network providers, third-party storage services, and third-party service providers. One or more of these nodes could sell or share Sara's data to other parties for analytic purposes. Figure \ref{fig:12}, shows Sara's thoughts of her data usages, and the actual usages of Sara's data.

Sleeping time, travelling habits, number of visitors, number of occupants in a specified area, and much other knowledge can be inferred from the security monitoring camera. It functions as an extra eye that is always watching and recording information. A similar real scenario has been raised in 2019, where Amazon Ring video doorbell announced that the videos recorded on their "Neighbours" app are used by at least 400 law enforcement agencies nationwide to help in criminal investigations \cite{crime1}. Ring video doorbell is an IoT device that is installed in front of an individual's property and continually detects motion and captures videos, offering users the ability to communicate via audio and video with the people passing by their property \cite{Ring}. The "Neighbours" app also provides users with real-time safety alerts from the local police department and the residents living in the same area \cite{Ring}. Although Ring's app grants the users the choice in opting out from sharing their videos with the authorities, there were many privacy concerns of the knowledge that can be inferred from the collected data, which can lead the police to obtain an official search warrant requesting individual's videos \cite{crime2}.

\begin{figure*}[t]
\centering
\includegraphics [ scale=.4]{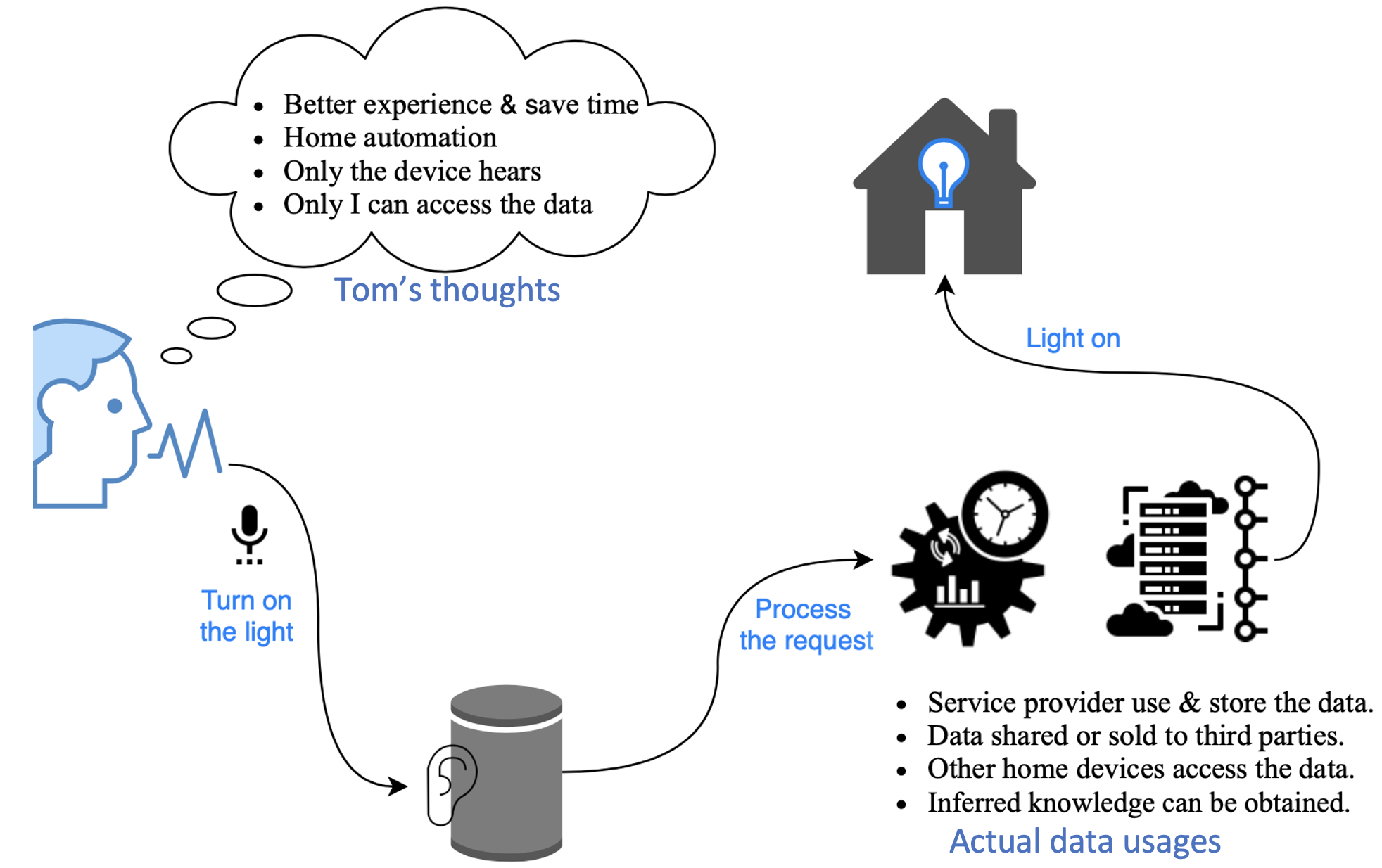}

\caption{Data usage use case scenario of a voice recognition system including the user expected data processing and the actual data processing.} \label{fig:11} 

\end{figure*}

2) Voice recognition systems: Tom works on a full-time job with changing shifts, takes care of his two children and volunteers in his town elderly day-care centre. To balance his daily activities and save time, Tom is using a voice assistant device that has a microphone, which can detect his voice commands and help him automate him home. Similar to the monitoring camera, the commands heard by the voice assistant system travel through many nodes in order for the device to perform its functionality. The commands are sent to a cloud-based system for processing, from which either a response is returned, or an action is performed on behalf of the user \cite{kumari2018multimedia}. The more voice commands the device hears, the smarter it becomes \cite{develop}, resulting in a massive collection of data that serve various purposes. For that, given the broad applicability of voice assistance systems, the data here are not only shared between the device and its required processing mechanisms; the data, however, is exchanged with multiple devices, which each has a different privacy policy \cite{hwang2015iot}. Figure \ref{fig:11}, shows a scenario of Tom's thoughts when he asks the voice assistance to turn on the light, and the actual usages of Tom's data.

Because voice interfaces are considered natural and do not require as much interaction as other interfaces \cite{moon2016voices}, they span many applications that have different sensing abilities \cite{hwang2015iot}. These applications, with their sensing feature, collect a considerable amount of data, which can lead to building a complete human profile, only through voice commands \cite{lau2018alexa}. The voice assistant device is somehow considered as an extra ear, which is always listening. In 2018, a case was raised by an Amazon Alexa customer, where their private conversation has been shared with others without their consent \cite{report4}. Moreover, according to the transparency reports released by Amazon, Apple, and Google \cite{report1}\cite{report2}\cite{report3}, law enforcement has sought data from 700,000 user accounts, which all have personal and sensitive information, and the companies have provided the information about two-third of the time. 

\subsection{Data Storage}
The common phenomena with IoT users, especially non-technical users, is that their data is safe and only stored in their owned device(s) \cite{zheng2018user}. However, with the technology development and the raise of high capability hardware and the Cloud service, the cost of storing data has dropped \cite{kumari2018multimedia}. Many organisations are moving towards storing their data in the Cloud \cite{zaslavsky2013sensing}. In the IoT context, the storage of the generated data is complicated. One IoT device might depend on multiple sensors to provide a service, in which each sensor requires different types and forms of data \cite{atlam2020iot}. The collected data will then be kept in the storage location(s) to be processed, and based on that; the IoT device will deliver the requested service \cite{zaslavsky2013sensing}.

\begin{figure*}[t]
\centering
\includegraphics [ scale=.4]{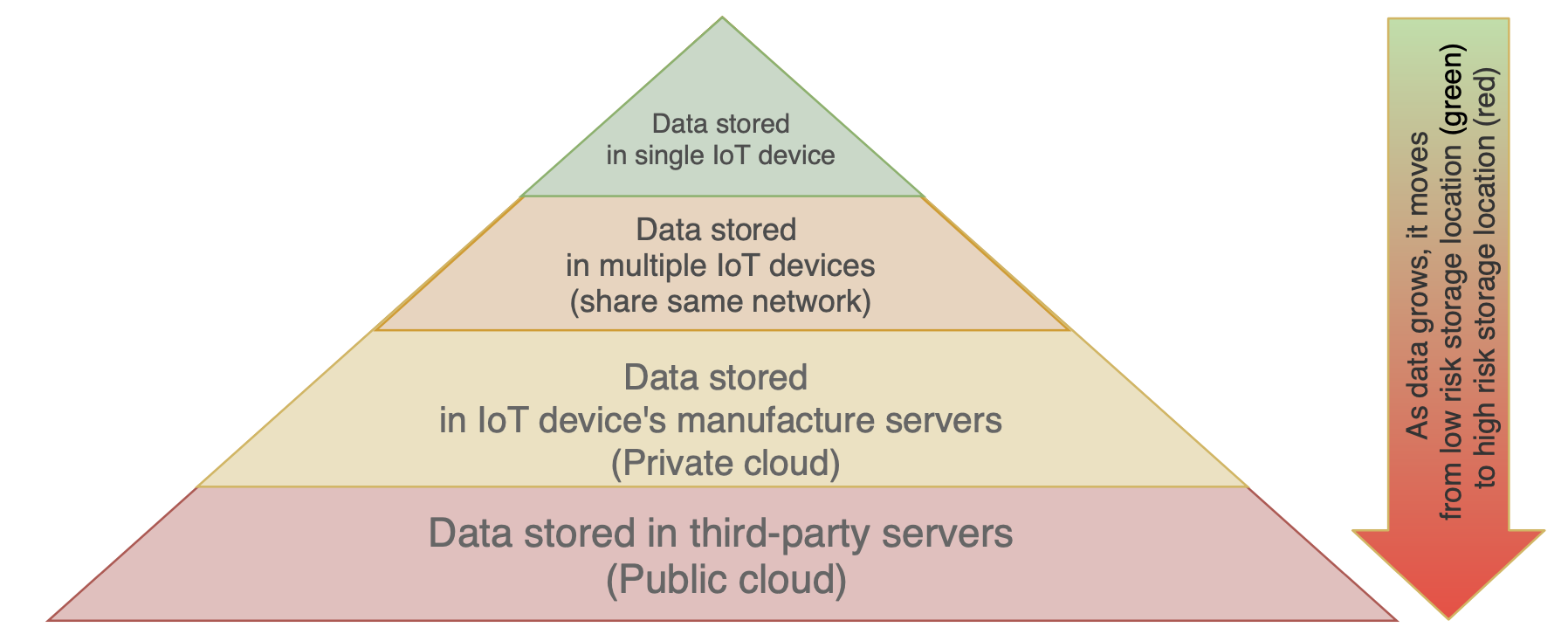}
\caption{IoT data various expected storage locations based on the device usage and originality. }

 \label{fig6:mesh6}
\end{figure*}

Storage locations and where the data is kept vary depending on the IoT device, the type of service it provides, and the producing manufacture. Figure \ref{fig6:mesh6} presents different data storage locations. First, the top layer is that the data is only stored within the device. Then, there is data that is stored within multiple devices in the same network. Getting out of the network boundary comes the data that is stored on the producing manufacture storage (private cloud). After that, is the data that is stored on third-party devices (public cloud). The last two differ in the location of the data, which might be within the country, or spans the entire world.

The feasibility and the efficiency of the internet-connected devices depend mainly on the collection of data that is done continuously without interruption \cite{atlam2020iot}. Given that, the IoT sensors will collect a high volume of data, which requires a computing power that cannot be handled by the small sensor. In addition to the sensor's collected data, storage space is also needed for the data analysis and annotation used to extract knowledge and patterns that are beneficial for the user \cite{atlam2020iot}. Hence, big data are produced, and Cloud services become an essential factor in providing data storage to IoT organisations \cite{zaslavsky2013sensing}\cite{reed2011imagining}. Medical, surveillance, energy, and many other data collected by IoT sensors are stored in the cloud \cite{kumari2018multimedia}\cite{apthorpe2017smart}. 

\subsection{Data Retention}
IoT builds its intelligence from data. An IoT device becomes better "smarter" as it collects and learns more data. Hence, devices that store data for an extended period or forever tend to function better than others \cite{develop1}. As depicted in Figure \ref{fig4:mesh4}, along with specifying the purpose of the data collection, websites' privacy policy usually mention the data retention. However, in the IoT domain, restricting the retention period is complex. That is due to the fact that has been mentioned earlier of IoT intelligence. Moreover, different sensors differ in memory size, application requirements, bandwidth, and throughput, which result in variation of the required retention period \cite{cao2019edge}. In order for the sensors to provide automation and ensure the user with a unique experience each time, it must learn from the user and compare with previous data \cite{atlam2020iot}. For instance, Amazon stated that Alexa voice interface gets smarter over time \cite{develop}, which means that the more Alexa listens, the better the service will be. Surprisingly, many users rejected the idea of retaining their data for an extended period by the various sensors \cite{chung2017finding}. Leon et al. \cite{chung2017finding} concluded that retention period plays a significant factor in the willingness of the users to share their data. Based on their study, users will less likely share their data when they are confronted that the retention period will exceed a week \cite{chung2017finding}.

Let us here lay down, how long data is kept by most IoT devices, especially devices that are used in the context of a shared space. Since IoT devices used inside the shared space usually has more than one sensor, the time period for data retention differ within one IoT device \cite{atlam2020iot}. An example of that is the smart thermostat, which adjusts the temperature, turns off the appliances, and sends alerts when it detects smoke \cite{lu2010smart}. Given that, it is clear that within the smart thermostat, there are at least three different sensors: temperature and humidity sensor, smoke sensor, and motion sensor \cite{lu2010smart}. Each one of these sensors stores data for a specific amount of time, which can be session-based, days, months, years, or an infinite period. In addition, theses sensors collect data on a regular basis, which can be every second, every hour, every day, etc. It is worth noting that, the retention period for most organisations tend to be over 12 months, which is the ideal time to perform analytic on the data, as well as acquiring other data \cite{develop2}\cite{McFadin}. The privacy policy of Nest thermostat, which is one of the smart thermostats that receives wide attention from buyers, declared the retention period for the collected data \cite{Nest}. They stated that some of the data are kept forever unless deleted by the user, while other data are kept for the period of thirteen months \cite{Nest}.   \\

\subsection{Notification Methods}
\label{Notification Methods}
As described in the introduction, the available notices fail in providing users with the appropriate IoT privacy knowledge. The other awareness factors will not be delivered to the users if the notification method fails. For this purpose, we provide an extensive survey of the notification methods. The goal is to present the different techniques available, which future research can build on. The importance of the notification mechanism derived from the extensive growth of IoT devices, where they have become deeply penetrated in everyday life in a way that made them unnoticeable by people \cite{edwards2005switching}. In 1991, Marc Weiser \cite{weiser1991computer} had described the "computer of the 21st century" as "The most profound technologies are those that disappear. They weave themselves into the fabric of everyday life until they are indistinguishable from it." Given the extent the IoT technology has reached, we can almost undoubtedly say that it represents Marc Weiser phrase.

\begin{figure*}[t]

\includegraphics [ scale=.45]{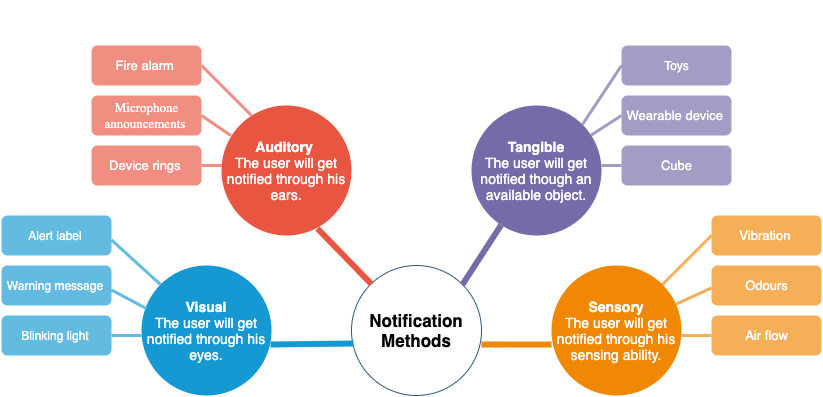}
\centering
\caption{Different types of notification methods, divided into four categories. Each category has various contexts of application.}

 \label{fig10:mesh10}
\end{figure*}

In the following section, we present several works of literature that have investigated notification methods in different contexts. We are categorising the research papers according to the notification methods they discuss (as shown in Figure \ref{fig10:mesh10}). As presented in Table \ref{table:4}, most of the notification methods fall into one of four categories: (a) Visual: including light, and motion notification methods. (b) Audio: including sound and motion notification methods. (c) Sensory: including vibrate, touch, and airflow notification methods. And (d) Tangible/physical: including wearable and cube format notification methods.

   \begin{table*}
   \small
   \centering
   \caption{Pieces of literature presenting the notification methods, the methods are divided into four categories, with each having different format.} 
\label {table:4} 
   \begin{tabular}{p{.21\textwidth}p{.039\textwidth}||p{.02\textwidth}p{.02\textwidth} p{.02\textwidth}p{.02\textwidth}p{.02\textwidth} p{.02\textwidth} p{.02\textwidth} p{.02\textwidth} p{.02\textwidth} p{.02\textwidth} p{1.5cm}}

\\\hline
                               & \multicolumn{10}{c}{Notification method}                                                                                                                                                                                                                                                                                                                                      &                                                                                                                                               \\ \hline
              &                 & \multicolumn{3}{c|}{Visual}                                                                                                                              & \multicolumn{2}{c|}{Auditory}                            & \multicolumn{3}{c|}{Sensory}                                                                 & \multicolumn{2}{c|}{Tangible}                             & \multicolumn{1}{c}{}                                                                                                                         \\ \hline
\multicolumn{1}{c}{\myrotcell{Author}} & \multicolumn{1}{c||}{\myrotcell{Citation}} & \multicolumn{1}{c}{\myrotcell{Motion}} & \multicolumn{1}{c}{\myrotcell{Light}} & \multicolumn{1}{c|}{\myrotcell{\begin{tabular}[c]{@{}c@{}}Visual\\(warning \\ notifications)\end{tabular}}} & \multicolumn{1}{c}{\myrotcell{Motion}} & \multicolumn{1}{c|}{\myrotcell{Audio}} & \multicolumn{1}{c}{\myrotcell{Air flow}} & \multicolumn{1}{c}{\myrotcell{Vibration}} & \multicolumn{1}{c|}{\myrotcell{Odours}} & \multicolumn{1}{c}{\myrotcell{Cube}} & \multicolumn{1}{c|}{\myrotcell{Wearable}} & \multicolumn{1}{c}{\begin{tabular}[c]{@{}c@{}}Control parameters \\ Intensity,  \\ frequency,\\ duration, \\modulation,\\ sequences \end{tabular}} \\ \hline

 \centering{Mehta, V., et al.}   &  \cite{mehta2016privacy}       & &   &   &   &   &   &   &  &   &  X & I, F, D, M, S \\ \hline
 \centering{Houben, S., et al.} & \cite{houben2016physikit}      &X & X &   & X &   & X & X &   & X &   & I, F, S \\ \hline
\centering{Kohanteb, O., et al.} & \cite{springrsensor}             & X & X & X &   & X &   &   &   &   &   & I, F, M\\ \hline
  \centering{Hornecker, E. and Buur, J.} & \cite{hornecker2006getting} &   &   &   &   &   &   &   &   &   & X &   \\ \hline
 \centering{Ishii, H. and Ullmer, B.} & \cite{ishii1997tangible}         & &   &   &   &   &   &   &   &   & X &   \\ \hline
 \centering {Jansen, Y., et al.}&  \cite{jansen2015opportunities}   & &   &   &   &   &   &   &   &   & X &   \\ \hline
\centering{Kohanteb, O., et al.} & \cite{summersensor}              & X & X & X &   & X &   &   &   &   &   & I, F, M \\ \hline
  \centering{Kubitza, T., et al.} & \cite{kubitza2016iot}            & & X & X &   & X &   &   &   &   &   & F \\ \hline
 \centering{Chernyshov, G., et al.}& \cite{chernyshov2016ambient}     &  &   &   &   & X &   &   &   &   &   & I, F, M \\ \hline
\centering{Haslgrübler, M., et al.}  & \cite{haslgrubler2017getting}    &&   &   &   & X &   &   &   &   &   & I, F \\ \hline
 \centering{Bodnar, A., et al.} & \cite{bodnar2004aroma}           &&   &   &   &   &   &   & X  &  &   &   \\ \hline
  \centering{Kaye, J.}& \cite{kaye2001symbolic}          &&   &   &   &   &   &   &   &   & X &   \\ \hline
\centering{Emsenhuber, B. and Ferscha, A.} & \cite{emsenhuber2009olfactory}   &   &   &   &   &   &   &   & X &   &   & I, M \\ \hline
 \centering{Olalere, I., et al.}& \cite{olalere2018remote}        &  &   &   &   &   &   & X &   &   &   & I, F, D \\ \hline
  \centering{Kumari, P., et al.}& \cite{kumari2015picam}         &  &   &   &   &   &   &   &   &   & X & I, F, D, S \\ \hline
 \centering{Corno, F., et al.}& \cite{corno2015context}        &   & X   &   &   & X &   & X  &   &   &   & I, F, S \\ \hline
  \centering{Pousman, Z. and Stasko, J.} & \cite{pousman2006taxonomy}    &   &   & X &   &   &   &   &   &   &   &   \\ \hline
\centering{Böhmer, M., et al.} & \cite{bohmer2014interrupted}    &  &   & X &   &   &   &   &   &   &   &   \\ \hline
 \centering{Leonidis, A., et al.} & \cite{leonidis2009alertme}     &X &   & X &   &   &   &   &   &   &   & I, F, M\\ \hline
\centering{Banerjee, S. and Mukherjee, D.} & \cite{banerjee2013towards}      &  &  &  X &   &   &   &   &   &   &   &   \\ \hline
 \centering{ Emami-Naeini, et al.}& \cite{emami2019exploring}      &   &   &  X &  &   &   &   &   &   &   &   \\ \hline
  \centering{Ardissono, L., et al.} &   \cite{ardissono2009managing}  &&   &   & X &   &   &   &   &   &   &   \\ \hline
  \centering{Simons, D. and Chabris, C.}& \cite{simons1999gorillas}        & &   &   &   & X &   &   &   &   &   &   \\ \hline
  \centering{Most, S., et al.} & \cite{most2005you}               &  & &   &   & X &   &   &   &   &   &   \\ \hline
  \centering{Wolpert, D., et al.} &  \cite{wolpert2011principles}     & & X &   &   & X &   &   &   &   &   & I, M \\ \hline
\centering{Emsenhuber, B.}  & \cite{emsenhuber2006integration} & &   &   &   &   &   &   & X &   &   & I, M \\ \hline
  \centering{Com, N} & \cite{com2008movie}              &  &   & &   &   &   &   & X &   &   &   \\ \hline
   \centering{Emsenhuber, B., et al.}&  \cite{emsenhuber2011scent}    && &   &   &   &   &   & X &   &   &   \\ \hline
\centering{Kowalski, R., et al.}   &  \cite{kowalski2013cubble}        & &   & &   &   &   &   &   & X &   &   \\ \hline
  \centering{Lin, S., et al.} & \cite{lin2011pub}                &  &   &   &  & &   &   &   &   & X & F, D, S\\ \hline
 \centering{Matscheko, M., et al.}  & \cite{matscheko2010tactor}       &  &   &  & &   &   &   &   &   & X &   \\ \hline

\end{tabular}  
\end{table*}


\subsubsection{Visual:}

 The notification methods in this category are the widest used methods \cite{houben2016physikit} due to their higher bandwidth, convenient setup, usage and access, and clear user delivery. Visual includes any notification technique that can deliver its message through the visible way, in which the receiver could understand the entire message only through his eyes. This is like alert labels, warning messages from websites, blinking light, and moving items, e.g., the movements of a swivelling camera.
 
Kohanteb O., et al. \cite{summersensor}\cite{springrsensor} have proposed an interesting notification method called 'Signifiers'. A signifier is an add-on feature that can be added to a device in order to deliver information to the user about the available active sensor. In their research, they have provided several visual notification techniques, which act as signifiers and can be implemented with minimal effort to guide the users' awareness of the sensors around them. They have tested their approach on eight IoT devices and their corresponding sensors. Their selection of the IoT devices was based on the devices that can have different usages within the house, different sensor modalities, and that can span to several domains. The main aspect of the research was to notify the users about the activity of the sensors around them in the least complex and annoyance method. For that, they have used a mechanism that represents what the device is collecting and that the average user could easily interpret. For instance, a flashing light on a camera can indicate that the camera is recording. In addition, pop-up signals on the sides of a device can indicate that the device is collecting audio data, i.e., the signals are placed on the device side to mimic human ears. Although their method can only inform the user about the data collection, it is easily understood and developed. The uniqueness of the information the signifier provides makes it readily adaptable by the developing manufactures. Their method of notification also has the feature of reducing the disruptiveness the user usually faces within different devices notifications \cite{pousman2006taxonomy}\cite{bohmer2014interrupted}, especially smartphone notifications. In \cite{bohmer2014interrupted}, the authors have discussed a method of showing a small notification at the top of the screen, which can reduce the user annoyance when he receives a call, and the notification abrupt the full phone screen.

Kubitza T., et al. \cite{kubitza2016iot} has presented an infrastructure that uses the ubiquitous nature of the IoT devices to deliver information to the users. They argued that the notifications must be delivered to the users in a context-sensitive and multi-modal way, reducing the need for smartphone usages. For that, they proposed a design that uses the meSchup IoT platform due to its intuitive setup and wide adaptability. Through the meSchup platform, users are able to get notified visually without the need to check their smartphone regularly. Whenever the smartphone receives a notification, it will be sensed through the meSchup notification gateway. It will be displayed to the user visually based on the available IoT sensor in the room. For instance, differences in an LED colour scale would indicate to the user a specific notification received on his smartphone. A more informative notification about the received notification could also be visualized to the user through, for example, smart TV and/or smartwatches. In order to protect the user privacy, the selection of the notification type is specified by the user depending on his criteria, and the availability of the sensors. Their method is similar to the approach presented in \cite{ardissono2009managing}, where they adopted a way that predicts the user's activities and based on that makes the decision of delivering, postponing or deleting the notification. This approach can effectively increase users' awareness about the surrounding sensors due to its unique sensor delivery modality. Here instead of the notification arriving at the user's smartphone, it will be forwarded to the most related sensor it is notifying about. As an example, a reminder about an appointment with the dermatologist could be displayed at the electronic mirror, which might get ignored when posted at the smartphone screen. The main limitation of this approach is the difficulty of its adoption in shared spaces, due to privacy risks it might trigger.

Using the visual approach to deliver notifications to the user plays an essential factor in increasing user awareness about his privacy. Various other studies and projects are using this type of notification, such as \cite{leonidis2009alertme}\cite{banerjee2013towards}.  When the user sees - through his eyes- a unique signal produced by a sensor, it will automatically lead him to take action regarding that sensor's which is collecting his information. The action that will be made by the user will depend on the level of awareness the user has about how the device is processing his data. The authors in \cite{emami2019exploring} presented a privacy label prototype, and they found that individuals purchase behaviour tends to change when they know the privacy implication of the IoT device.

\subsubsection{Auditory:}
This category includes the notification methods that can be heard, in which the receiver will get notified through his ears. Similar to the visual techniques, this method has higher bandwidth which is also considered a popular notification technique \cite{houben2016physikit}. Examples of audio notification are fire alarm, microphone announcements, mobile phones rings and the sound of the moving items, e.g., camera shutters.

Chernyshov G., et al. \cite{chernyshov2016ambient}, presented a novel audio notification approach. The approach proposed by the authors can help average users understand the status of the IoT device with no previous knowledge about the meaning of the audio notification. They used melodic rhythm to deliver information so that the user can perceive the information in an interesting hand-free and eye-free way. In order for straightforward interpretation, the sound samples they used in the melodic rhythm are associated with the process it is representing, i.e., they recorded the sound of the printer to represent printing. The rhythmic method used in this paper has many significant advantages. It does not only notify users about the active IoT device, but it also provides a continuous notification about its statutes, which is useful especially in the IoT domain since users might forget about the existence of the device after a while. Another advantage is that this approach uses rhythms, which can convey more information to the user in a less obtrusive way when compared with the discrete sound notifications. However, having all the ambient audio notifications to be delivered to the user in one melodic rhythm might cause confusion and difficulty in distinguishing the type of device generating the specific effect, which is discussed in \cite{corno2015context}. In \cite{corno2015context}, the authors used a machine-learning algorithm to manage the notification based on the context and the user habit. Their system design has the ability to decide the person receiving the notification, the device, the perfect time and the ideal mode.

Haslgrubler M., et al. \cite{haslgrubler2017getting}, described a set of different notification methods that can be used in an industrial environment. Their purpose was to develop the best approach that can direct and alert the industrial workers about potentially harmful situations. Apart from the visual and haptic methods they have proposed, they described the effectiveness of the auditory notification, especially in an area with workers of different background. Because of the environment of the research, i.e., industrial environment, they have used stationary speakers to send audio notifications. The sensors in the speakers will send warning sounds whenever the working machines reach a specified level of danger. It is also worth noting here, that the type of the audio notification delivery can differ based on the environment, in a similar environment, earplugs audio notifications, for example, will not be as effective as the stationary speakers as they might withdraw the user attention of his surroundings. The drawback of this technique is as described by several other studies that even if the notification is delivered in the correct time and modality, it might get unnoticed \cite{simons1999gorillas}\cite{most2005you}, leading to the lack of attention that might occur in an environment with a loud noise. The preceding, however, can be mitigated with only delivering the notification that is task-relevant \cite{wolpert2011principles} and increasing the notification intensity. 

The audio notifications offer a more natural way of communication with the user. Users are easily notified, even when they are busy with other tasks with minimal to no interruption as opposed to other methods like the visual method, which requires eye contact. Many other pieces of research presented audio notification as well, where some papers we have discussed in our survey. As an example, the sounds generated by the camera is considered as an audio notification method in  \cite{summersensor}\cite{springrsensor}, e.g., the sound of shutter opening and closure and camera swivels. The sound of the push notifications in the smartphones and Amazon Alexa is perceived as an audio notification method in \cite{kubitza2016iot}.

\subsubsection{Sensory:}

The sensory notification is a method that sends the information to the user through various sensing mechanism, and the user as well receives the information through his sensing ability. It includes touch, smell, feel and taste to send and receive the information. For instance, mobile phones vibration is considered a sensory notification method. The smell and feel of the smoke are also considered sensory notification methods. This category is not widely used like the visual and auditory notifications since it is hard to set and interpret \cite{bodnar2004aroma}\cite{kaye2001symbolic}.

Olfactory notification approach has been discussed in several pieces of literature \cite{emsenhuber2009olfactory}\cite{emsenhuber2006integration}\cite{com2008movie}\cite{emsenhuber2011scent}. Emsenhuber and Ferscha \cite{emsenhuber2009olfactory} proposed the olfactory interaction zones (OIZs) as an effective mean of communication. They discussed that the odours emitted either by humans or other entities convey information, which can be detected through the available sensors, e.g., gas sensors or electronic nose. This process of detection and processing the olfactory information is what they referred to as OIZs. The OIZs, as they proposed, provides a spontaneous interaction that can be identified by people and machines. They argued that although odours may be hard to be applied and volatilise fast, they can be easily recognised and usually refer to the specific situation of data. This feature distinguishes the olfactory method, as it provides in-depth information to the user comparing it with the other techniques, which provides the user with abstract information only.

The sensory notification mechanism can also be used to send fault-tolerance alerts, in which it notifies the users about a potential fault in the used system. Olalera I. et al. \cite{olalere2018remote}, proposed the use of remote condition monitoring (RCM) as a method that can support proactive machine maintenance through vibration notification. Whenever there is a fault in the machine, it will send a vibration signal, and based on the severity of the vibrations, the users will be notified. Using this approach has proved the ability to proactively detect malfunctions in the system and respond respectively. While the vibration method is simple to adopt, it can provide a unique notification way to the user. In another paper \cite{houben2016physikit}, for instance, the vibration mechanism has been used as a notification method, even after the vibration has stopped. They examined the use of an object, e.g., a plant, which can be placed over a vibrated cube, in which based on the direction of the object, the user will understand the desired notification. 

The sensory notification method provides a unique tactic of notifying the user. Its characteristics make it deployable to span a wide range of people, including people with special needs \cite{kumari2015picam}. It also delivers a timely notification more smoothly and intuitively, eliminating the disruptiveness the user might encounter with other the notification methods.

\subsubsection{Tangible/physical:}

One of the new notification techniques is the tangible or physical method. It is when the user receives information though an available object. An example of that is the wearable Fitbit watches, in which they could send sensory information such as heat when overused. Although it is using a sensory notification, it lies in this category due to the fact that the tangible object must exist in order for the information to be delivered to the user. This approach is new but growing, where it does not substitute the available notification methods, but instead provides a hybrid way of notifying the user. 

A 'human data design' approach was proposed by Houben et al. \cite{houben2016physikit}, where they developed what they called a Physikit. The Physikit is a toolkit and technology probe which uses several physical cubes called PhysiCubes. It allows the users to receive notifications about the usage of their data in a physical, tangible format. The Physikit requires two elements to perform. First, several tangible physical cubes, which each deliver one unique notification, such as movement, light, air or vibration. Second, a web-based end-user configuration tool, which provides the user with an easy way to connect his data sources. Through developing this Physikit, the authors argued that when tangible physical objects are made available to users, it will give them the urge to explore more about the collection and processing of their data. As a result, the users will have the confidence to make thoughtful decisions regarding the share of their data. In their research, they have implemented the Physikit and conducted a field study to assess the usability of the tangible physical notification method. The overall results were satisfying, in which users - households mainly - of different background showed positive engagements with the physical cubes. The users’ data awareness has improved in a way that some users were interested to know how to set rules to manage their data. Despite that the Physikit provided a powerful technique in spreading the awareness, it, unfortunately, holds some limitations. Since users are given a choice to set up the information the cube notifies them about, it creates a hurdle in setting the rule, understands the rule, and memorise the information the cube is trying to notify about. It also led to conflicts of interests, since each house member could set the cube to notify him about a different data change. Furthermore, the approach being connected to a web-based technique puts a limitation in front of non-technical people (e.g., elderly) trying to set it up. The cube idea has also been used in \cite{kowalski2013cubble}, as an effective way of notifications. In this paper, the cube has been used as a mean of communication to people in a long-distance relationship. The cube has similar functionality to the Physikit cube in which it shows light, vibrates or heats up whenever to notify the users. 

Another tangible notification technique was presented by Mehta et al. \cite{mehta2016privacy}, in which they explored the efficiency of the on-body notification methods. The authors argued that using on-body haptic interfaces could provide the user with awareness regarding the use of his data while preserving his privacy. They presented two main functionalities using the metaphors' privacy itch and privacy scratch’ in their proposed wearable tangible device. (1) Privacy itch: which causes an itch in the user's arm to warn him about potential personal data breaches, and (2) Privacy scratch: which allows the user to scratch his arm as a response to the itch, providing a real-time, contentious and eye free control of his privacy preferences. The on-body privacy management notification method offers a practical and useful way of real-time notification. Its ease of use makes it span not only to technical people but also to users with different backgrounds providing them with convenient interaction way with their data. In addition to that, the on-body notification provides the users with distinct interaction feature, which gives them trust through having the dominance in controlling their data. However, this technique could convey only a limited number of information, resulting in the user uncertainty when dealing with his private data which either will lead to ignorance or great concern. Moreover, since this notification method is attached to the body, it can be really obstructive to users as described by \cite{lin2011pub}\cite{matscheko2010tactor} especially when multiple warning is sent in a limited amount of time.

Considering the above papers and the studies in \cite{hornecker2006getting}\cite{ishii1997tangible}\cite{jansen2015opportunities}, we can see that tangible and physical notifications can - with some improvements - be a promising method in increasing the users' awareness. Their physical feature makes them at the sight of the user's eye most of the time, triggering the user's curiosity in learning about his data. In addition, based on the tangible interface, it can provide the user with various location properties unlike the usual notification methods, i.e., they can be moved between different rooms in the house, or shared between different members.\\

\subsubsection{Control Parameters}

With the various notification methods available, each and every method could convey different information based on how it is controlled. The visual light notification, for example, can have a different intensity to indicate the severity of the delivered notification, e.g., strong red represents very sensitive data. It can also have different colours, in which each colour represents a particular data type, e.g., green represents normal generic data, i.e., data that is generally accepted to be shared. There are mainly five different ways that can play as a control parameter within the available notification methods. They are intensity, frequency, duration, modulation and sequences. The five control parameters could be applied mostly to every notification method to deliver a wide diverse range of notifications to the user. Table \ref{table:4} presents the papers that incorporate the different control parameters in their used notification criteria. \\

\subsubsection{Notification-based interaction}
Understanding user behaviour is an essential piece in choosing or designing the appropriate notification technique. Based on individuals' behaviour, reactions, and interactions with a particular device, the privacy policy can be adjusted and modified. Several pieces of literature have confirmed the importance of human-factor in supporting the setting of the privacy policies and aspects \cite{baarslag2017automated}\cite{patil2015interrupt}. Furthermore, different researchers have discussed the importance of embedding the feedback feature while developing a device that interacts with individuals. In this section, we extend on including some of the research papers that discussed allowing the users to interact and submit feedback based on the notification he/she receives from a particular used device.

Most devices, especially IoT sensors, if incorporating an interaction mechanism with the user are employing a mobile application to send notifications and receive feedback to/from the users \cite{koreshoff2013internet}. Gordon et al. \cite{gordon2016participatory}, have designed an interactive application that can track users health. The application allows the health provider to send notifications, and respectively allow the patients to specify what information they want to share with their health provider. Lee et al. \cite{lee2018iot} and Kiljander et al. \cite{kiljander2014semantic} discussed similar contribution in their papers. The development of these applications provide the users with interaction with their devices which allow them to set their privacy choices, however, it usually gets neglected by the users due to their setting requirements difficulty. On-body interactions, in which a person performs a body action or movement like smiling or blinking, have also been used as a feedback option to assist privacy choices \cite{matscheko2010tactor}\cite{mehta2016privacy}\cite{sherrick1991vibrotactile}. In \cite{mehta2016privacy}, Mehta et al. presented a privacy band that uses an on-body haptic interaction to send notifications to the user. Based on the received notification, the band allows the user to replay to the notification by submitting feedback which includes his/her privacy preferences. The on-body interactions have shown its promise in preserving individuals privacy, but it suffers from the annoyance the users incorporate from multiple notifications.

\section{Discussion}
\label{Discussion}
Based on the literature that we have reviewed, and as stated in \cite{corno2015context}, the available notification models do not cover the entire privacy image the user needs to understand. For instance, some methods only inform the user about the operation of a sensor in the room but do not specify the purpose of the sensor operation neither specify the length of the period the sensor will operate \cite{poppinga2014sensor}. With the massive growth of the IoT gadgets, it is alarming how these sensors can sweep data without the user's knowledge \cite{calo2011against}. The usage of the collected data certainly provides user benefit, but what is unknown is that the large volume of the collected data could make a personal data market which is created through users' trust \cite{tragos2016trusted}. Manufactures usually obtain their customers' consent regarding the collection of their data \cite{tzafestas2018ethics}\cite{weber2015internet}; however, a considerable gap and trade-off are facing the customers' perception. Often customers feel hopeless when it comes to the possession of an IoT device. They feel lost in front of the long privacy policies, leading them to provide their consent only to be able to acquire the benefit of the IoT device \cite{steinfeld2016agree}.

Most users are not aware that by providing their consent, their data is usually collected in a large amount \cite{steinfeld2016agree}. That is because each IoT device has more than one sensor that is collecting data on a regular basis. With that happening, the volume of the collected data proliferates beyond the ability of the manufactures' servers' storage, leading to the necessity of Cloud involvement. When the cloud comes into the picture, it opens the horizon to third-party companies to have access to customers' data. Average users might trust their IoT device company, but not the third-party. Third-party companies, in this case, have access to data, which might contain personal information, and would use them for various purposes under the cover of already obtained user consent. Che et al. \cite{che2013big}, and Tene et al. \cite{tene2011privacy} presented several privacy concerns, such as profiling, stealing and targeted ads that have been raised, arguing that users confidentiality has been revoked. In addition, user's data are not only accessed by third-party, but data is also retained and/or archived for a period of time, which sometimes can be infinite. According to \cite{naeini2017privacy}, when people were confronted with the location of their data and how long the data is kept, most of them show preferences to devices that either offer a short retention period or an option of data deletion. Furthermore, several studies like \cite{barua2013viewing}\cite{lee2016understanding}\cite{lee2017privacy} show how individuals care about their privacy and demand the possession of the data, which is collected based on their habits and behaviour. 

\begin{figure*}[t]
\centering
\includegraphics [scale=.5]{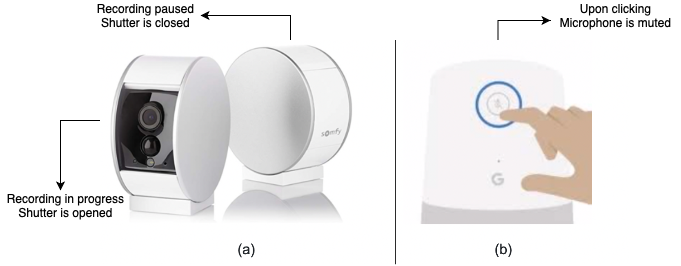}

\caption{Two IoT products that incorporate special features, which can support in preserving individual's privacy. } \label{fig7:mesh7}

\end{figure*}

There are some new solutions that have been produced in the market to support preserving the privacy of the user. Somfy, shown in Figure \ref{fig7:mesh7}(a), created a monitoring camera with a privacy shutter, where the shutter closes whenever a person enters his private area \cite{Somfy}. They have guaranteed that if the shutter is closed, nothing is recorded or stored in the cloud \cite{Somfy}. Google has a smart speaker, shown in Figure \ref{fig7:mesh7}(b), with a physical microphone switch that can be turned on and off, according to the user desire \cite{Google}. It can be challenging to assume that all IoT devices can involve a privacy feature since each device is equipped with different sensors that are collecting different information. However, it is essential to have a common phenomenon of preserving user privacy and informing the users about any related mean performing data collection. IoT devices manufactures should work with the application developers to satisfy the end users' privacy needs.

IoT developers and manufacturers need to consider the privacy aspects of the device and how it could impact the customers throughout the entire device development cycle. Sensors in the connected world have introduced new ways of data collection, which, as a result, bring multiple privacy challenges. Adopting the P3P protocol ontology, as described earlier, could help in addressing some of these challenges. For instance, granting the user the choice to control his data was one of the essential features of the P3P protocol \cite{p3p2}. Although the P3P was a web-based tool, its mechanism can help in reducing the privacy issues faced in the IoT domain \cite{ghazinour2009lattice}\cite{langheinrich2002privacy}. Another effort has also considered the privacy challenges and presented a way of allowing the users to specify the location, the duration, and the kind of data they wish to be stored within the company cloud service has been discussed in \cite{satyanarayanan2015edge}.

Perhaps the most significant challenge regarding the users’ IoT privacy is how to get the users to absorb and understand the sensitivity of the operations and the processes done on their data. Especially for non-technical users, classic visualisation, i.e., similar to usual privacy policies, carry little to no meaning to the users about their data collection and usage \cite{balestrini2014beyond}\cite{balestrini2015iot}. Current research focuses on developing a notification method that is easy to understand and does not disrupt the user while simultaneously enhancing user awareness \cite{emami2019exploring}. The more aware the user is, the better the privacy policy can be built, which serves both the users and the IoT developers \cite{sadeh2009understanding}.

\section{Research Challenges and Opportunities}
\label{Research Challenges and Opportunities}
With the new growing modes of interactions introduced through the IoT devices, challenges of people understanding the underlying idea of these interactions arise \cite{steinberg2014these}\cite{fruchter2018consumer}\cite{curran2018your}. The notification mechanism is employed extensively in the digital space to notify users about how websites are using their interactions \cite{kelley2009nutrition}. For instance, users surfing the web, usually, get informed that their data is being collected and used to provide them with the service. Besides, in most cases, users are deliberately accessing web services, with their choice and knowledge. However, this is different in the IoT context \cite{edwards2016privacy}. The IoT devices and sensors are widely adoptable, implemented in physical spaces, and are considered small in size, creating them the perfect environment to go un-noticed \cite{kelley2009nutrition}.

In this section, we elaborate on the gaps discussed previously, and present some of the research challenges and opportunities for future research. In Section (\ref{Privacy Infrastructure}) we discuss whether having a formal notification infrastructure can help in increasing the users awareness? If yes, what development language should be used to develop the infrastructure (section \ref{Development Language})? Furthermore, there is a lack in having a unified interaction patterns between the user and the IoT device, which is discussed in section (\ref{Interaction Patterns and Personalisation}). We then highlight additional IoT interaction challenges that can offer new research opportunities (section \ref{Additional challenges}).

\subsection {Privacy Infrastructure}
\label{Privacy Infrastructure}
Unlike the websites in digital spaces, which the person utilise with his/her choice, the collection of data done through IoT devices in physical spaces is often unbeknown to the people \cite{van2017better}\cite{gerber2018foxit}. Although, after the enforcement of the GDPR and the CCPA regulations, most public spaces under surveillance, such as banks, universities, shopping malls, etc., are employing some notification methods, this appears to be not sufficient. The notification method that is often used is usually a warning sign indicating that there is, for example, a camera in progress, with no further information. These signs, despite their benefit, convey little to no meaning about what is done with the collected data. Additionally, in most cases, people will forget about their existence \cite{erickson1995years}\cite{gutwin11amp}. There is a clear lack in the availability of the resources that inform people about the surrounding technology that collect and process their information. 

A novel contribution by \cite{das2018personalized}, where they designed an IoTPI app. The app will inform the users about the registered IoT devices available within their vicinity. Based on the registered devices policy, the user is able to browse the processes done on his/her data, as well as, some devices will offer the option of opt-in or opt-out. A research question that can be raised here, is whether a similar approach can be followed to span a wider audience. How can a person be notified and reminded about the existence of an IoT device without the need for a smartphone? Additionally, how can an average person understand the collected data type, how it is used, stored, or how long it is retained? Can the IoT device itself incorporate a notification mechanism? Or it has to be done through an external device that can comprise more than one IoT devices, i.e., devices that use each other?

Augmenting existing IoT devices, in which one IoT device can use other IoT devices' capabilities, has been introduced in \cite{angelini2018internet}\cite{pokric2014augmented}\cite{guo2013internet}. An example of that is, a touch screen on a smart fridge can display the settings of a nearby coffee machine, allowing the user to acquire more than one service at the same time. Although this technique has shown its advantages in aiding energy-saving and minimising the time-spent while using an IoT device, individuals' privacy can get compromised. The devices sharing capabilities drive at having the user's consent to exchange his/her data between two or more IoT devices. Doing so indicates that the user's data can be processed and stored at different manufactures' storage without the users' knowledge, where some might not be following strict privacy policies. Having a privacy infrastructure that conveys to the user the processes done on his data in similar scenarios will help in increasing the user's privacy awareness.

\subsection{Development Language}
\label{Development Language}
A major challenge that arises with the IoT emergence is the diverse nature of its developers. IoT devices in the market are not only developed by known reputable companies that have access to resources, but also are developed by small entities or individuals that may lack essential resources and/or experience. Consequently, IoT devices are acquired and used by almost all levels of society. So, in order for having an IoT device that supports the user's privacy, it is essential to employ a development language that is fast and reliable. More importantly, there is a need for a development language that contains the privacy required information to serve both the developer and the device user.

As described earlier, the P3P protocol gives the user control regarding the use of his data. The inclusion of P3P into the IoT domain as a mean that can increase individuals' awareness has been proposed by Langheinrich \cite{langheinrich2002privacy}. Langheinrich proposed a model that uses the P3P machine-readable privacy policies to communicate with nearby IoT sensors, allowing the users to manage their preferences regarding their personal information. Ghazinour et. al \cite{ghazinour2009lattice}, have built upon the use of P3P in presenting a model that does not only provide the privacy policy to the user but also ensure the enforcement of the use of the privacy policy by both the user and the service provider. Other languages, such as EPAL \cite{ashley2003enterprise} and PPVM \cite{ghazinour2009model} have also incorporated privacy policies that can support in the IoT domain. Although with the IoT sensors, there is a considerable amount of sensed data, it is practical to have the employment of the P3P protocol, due to the fact that the enforcement of policy is usually task-based.

There are various development languages that are available for developers, such as \cite{arnold2000java}\cite{wilde2011swift}\cite{kernighan1988c} \cite{van2007python}. Although these languages are powerful, they mostly require web-based tools and are directed to individuals with a technical background \cite{naeini2017privacy}. Considering that the IoT devices, specially the small and unnoticeable devices, are developed by individuals and small entities, the privacy requirements are usually get neglected due to their complexity, cost, and building difficulties. Moreover, the developers of these devices have a little experience of employing any device privacy updates \cite{hong2017privacy}. Will having a development language that is reliable, and cost-effective help in incorporating the privacy requirements into the IoT devices? Or can there be templates that is needed to be followed in order for a device to pass a privacy check? Will third-party involvement help in tackling this issue? 

\subsection{Interaction Patterns and Personalisation} 
\label{Interaction Patterns and Personalisation}
Considering the diversity of the IoT devices, devices that lie in the bottom-tier layer \cite{hong2017privacy}, are those that usually fall from a person's attention. That is due to their size and multiplicity, where an individual can have hundreds of them \cite{hong2017privacy}. Having this great number of devices around a person rise the privacy issue of the amount of data that is collected through them. In particular, these devices have low awareness models while collecting individual data, making them a threat to the person's privacy \cite{niemantsverdriet2019designing}\cite{neustaedter2006linc}\cite{crabtree2004domestic}.

To elevate the devices' awareness models, an interaction pattern between the individual and the IoT device is needed. This interaction pattern should be reliable and effectively communicate the flow of data to and from the device \cite{langheinrich2002privacy}. Moreover, the interaction pattern must be readily deployable considering the size of the sensors and the experience of the device user. Can there be an interaction pattern that conveys to the user an essential device's functionality in a straightforward way, e.g., red blinking light indicating sensitive data collection or loud sound indicating an urgent needed interaction? Will the interaction pattern cope with the number of sensors acquired by a single individual? Can we have a cost-effective model that balances the number of needed notification with user's annoyance? 

The availability of such an interaction pattern requires an understanding of both the user and the device. In the case of IoT, a comprehensive understanding of the users' social context and the IoT sensor functionality is a must. That is because, the IoT sensors are shared in nature, i.e., either they are deployed in a shared space or are used by more than one person. There exist multiple designs and frameworks that support understanding individuals' awareness level, such as \cite{niemantsverdriet2019designing}\cite{chung2004development}\cite{iachello2008privacy}\cite{arora2019design}\cite{zimmerman2009designing}\cite{borchers2008pattern}. However, most of these frameworks are situated to target experienced developers and users, making them difficult to be adopted in the IoT domain, since a great amount of IoT sensors are developed and used by individuals or small entities. In addition, the available frameworks and designs are difficult to operationalise in IoT shared spaces. There is a persistent need for a unified interaction pattern toolkit that serves both the developers and the device users. We suggest having a toolkit for IoT interaction patterns since it will simplify the privacy awareness check for both the developers and the IoT device user. The toolkit can serve as a catalogue that includes different types of privacy interaction patterns, each with its advantages and disadvantages, where the developer and/or the device user can create or choose from the recommended personalised patterns according to their needs. In addition, the toolkit can also be adjusted to cope with the number of IoT sensors occupied in one shared space, and the number of the notifications they arise. Having a framework or a toolkit that is easily adoptable and deployable will serve in setting the first stone for the developer in taking into consideration the individuals' privacy while developing an IoT device. It will also give the IoT users' control over their data, which will increase their privacy awareness.

\subsection{Additional challenges}
\label{Additional challenges}

In this section we present an overview of two more challenges that can be studied by researchers in term of IoT sensors. First, is whether employing the notification methods in the IoT sensors can add up to increasing the individual's awareness and affect their decision making? Specially, will a person be nudged by the notification method and adjusts or alters his habits due to having a sensor in the room? Will the notifications have the same level of influence on all users or will there be different scale of influence depending on the user age and personality? Second, can there be a unified approach to deliver the notifications to the user? Will having a unified approach help in reducing the notification annoyance? Or can this approach help users in understanding the type of notification being presented \cite{lin2012expectation}? And like the previous challenge, will a unified approach fits all types of users despite their different age, personality, and life style?

\section{Conclusion}
\label{Conclusion}

In this survey, we have reviewed a number of the available literature that address different mechanisms of user’s notifications. The goal is to provide an in-depth study, which can help in improving the internet-connected devices users’ awareness. For that, we have classified the available notification methods into four main categories: visual, auditory, sensory, and tangible or physical notifications, along with providing a look into the pieces of literature that proposed the possibility for the user in replying to the raised privacy notifications.Furthermore, we have provided a look at the literature that discussed the most critical factors that should be taken into consideration while developing an IoT notification method. These factors are the collected data type, the purpose of data collection, the data storage location, and the data retention period. A number of gaps and challenges have been identified along with the survey, as well as recommending some opportunities and schemes which can serve as future research questions and help in addressing the suggested gaps.

\maketitle
\newpage

\newpage

\bibliographystyle{unsrt}  
\bibliography{library}
\end{document}